\documentclass{sig-alternate}

\usepackage{amssymb}
\usepackage{graphicx}
\usepackage{amsmath}
\usepackage{subfigure}

\renewcommand{\P}{\ensuremath{\mathbb{P}}}
\newcommand{\E}{\ensuremath{\mathbb{E}}}

\begin{document}

\title{Collective Attention and the Dynamics of Group Deals}
\numberofauthors{5}
\author{
% You can go ahead and credit any number of authors here,
% e.g. one 'row of three' or two rows (consisting of one row of three
% and a second row of one, two or three).
%
% The command \alignauthor (no curly braces needed) should
% precede each author name, affiliation/snail-mail address and
% e-mail address. Additionally, tag each line of
% affiliation/address with \affaddr, and tag the
% e-mail address with \email.
%
% 1st. author
\alignauthor
Mao Ye\titlenote{Mao Ye is also a PhD student in the Department of Computer Science and Engineering, the Pennsylvania State University, Pennsylvania, USA}\\
       \affaddr{Social Computing Group}\\
       \affaddr{HP Labs}\\
       \affaddr{California, USA}\\
       \email{\small mxy177@cse.psu.edu}
% 2nd. author
\alignauthor
Chunyan Wang\\
       \affaddr{Dept. of Applied Physics}\\
       \affaddr{Stanford University}\\
       \affaddr{California, USA}\\
       \email{\small chunyan@stanford.edu}
% 3rd. author
\alignauthor Christina Aperjis\\
       \affaddr{Social Computing Group}\\
       \affaddr{HP Labs}\\
       \affaddr{California, USA}\\
       \email{\small christina.aperjis@hp.com,}
\and  % use '\and' if you need 'another row' of author names
% 4th. author
\alignauthor Bernardo A. Huberman\\
      \affaddr{Social Computing Group}\\
       \affaddr{HP Labs}\\
       \affaddr{California, USA}\\
       \email{\small bernardo.huberman@hp.com}
% 5th. author
\alignauthor Thomas Sandholm\\
       \affaddr{Social Computing Group}\\
       \affaddr{HP Labs}\\
       \affaddr{California, USA}\\
       \email{\small thomas.e.sandholm@hp.com}
}

\date{}

\maketitle

\begin{abstract}
We present a study of the group purchasing behavior
of daily deals in Groupon and LivingSocial and introduce a predictive dynamic model of collective
attention for group buying behavior.
In our model, the aggregate number of
purchases at a given time comprises two types of processes: random
discovery and social propagation. These processes are
very clearly separated by an inflection point. 
Using large data sets from both Groupon  and LivingSocial we show how the model is able to predict the success of group deals as a function of time.
We find that Groupon deals are easier to predict accurately earlier in the deal lifecycle than
LivingSocial deals due to the final number of deal purchases saturating quicker.
One possible explanation for this is that the incentive to socially propagate a deal is based on
an individual threshold in LivingSocial whereas it in Groupon is based on a collective threshold, which
is reached very early. Furthermore, the personal benefit of propagating a deal is also greater
in LivingSocial.
\end{abstract}

% A category with the (minimum) three required fields % to be revised




\section{Introduction}
Attracting the attention of potential customers in today's
information rich social media is a challenge. As a result marketers
have been forced to target customers in more sophisticated ways.
Location-based (regional) and hyper-location-based (within
eye-sight) targeting has turned out to be very effective in terms of
improving conversion rates from views to
purchases~\cite{jiwire}. However, since people are unwilling to
share their exact locations out of privacy concerns they need to
be given some incentive to reveal their position. The most
successful incentive employed to date is daily
deals.\footnote{http://www.bynd.com/2011/05/04/social-loco-research/}
In spite of the success of this strategy it is not fully understood what makes it successful
and what kind of social behavior the daily deals sites so effectively tap
into and exploit. However, it is clear that deadlines and social propagation
play important roles in addition to location-based targeting. 
The main question we are addressing in this work is how to describe the purchasing pattern 
more precisely in order to predict the future popularity of a deal.

%first use renewal theory to model the random discovery process.
%After tipping, we use a
%multiplicative process to describe the social propagation behavior and
%explain the purchase growth behavior seen both on Groupon, the current market
%leader of daily deals in the US~\footnote{http://groupon.com} and LivingSocial.
We analyzed data from Groupon and LivingSocial, the current market leaders of daily deals in the US. 
Groupon promotes deals for different geographic markets, or cities,
called divisions. In each division, there is typically one featured
daily deal.  A deal is a coupon for some product or service at a
substantial discount off the regular price. Deals may be available
for one or more days. Coupons are only redeemable if a certain
minimum number of customers purchases the deal, and this number constitutes what Groupon calls a {\it
tipping point}. Furthermore, sellers may set a maximum threshold
size to limit the number of coupons that can be purchased. 
LivingSocial is similar to Groupon, except that there is no tipping point. The incentive that drives users to buy deals is the following commitment made by LivingSocial: ``Buy first, then share a special link with friends, if three friends buy, yours is free!". \footnote{http://www.livingsocial.com} %So it will be of great interests to see how the purchase dynamics changes in LivingSocial against Groupon. 

A closer examination of the mechanisms driving user behavior in
group deals could provide useful guidance for local marketing
campaigns. In this paper we study the evolution of
collective attention measured as deal purchases.  We base our
analysis on data collected from Groupon over two months and from LivingSocial over one month. Our
assumption is that successful deals arise from two
behavioral processes: random discovery; resulting from the
serendipitous discovery of a deal on the web portal, or in the
mobile app, or via an email subscription; and social propagation;
which results from the propagation of deals over social networks. These
processes are separated by an inflection point, which in Groupon is
the tipping point, after which there are enough purchases to
guarantee deal transactions. Before the inflection point is reached the customer
base is small so the random discovery process dominates. Conversely,
after the inflection point, a critical mass of customers have discovered
the deal to make social propagation dominate the purchasing
behavior.

The contributions of this paper fall into two categories:
\begin{itemize}
\item{{\bf Structure of purchasing dynamics.} We present a stochastic model that analytically explains the observed purchasing behavior.}
\item{{\bf Prediction model for purchases.} We show how the model is able to predict the success of group deals as a function of time.}
\end{itemize}

The paper is structured as follows. In Section~\ref{sec:related}, we discuss related
work.  In Section~\ref{sec:data}, we discuss the data sets and the collection
strategies used in our study. Section~\ref{sec:model} describes our stochastic
model and verify it empirically. Then in Section~\ref{sec:prediction} we use
our model to predict purchase volume and benchmark it against 
some baselines. 
%In Section~\ref{sec:comparison} 
%we compare the performance of deals in Groupon versus 
%LivingSocial both from the perspective of a deal site and a merchant.
Section~\ref{sec:conclusion} concludes with possible
applications of our work and future directions.

\section{Related Work}\label{sec:related}
The related work comes from two broad areas, social purchasing
behavior, and collective attention.

\subsection{Social Purchasing Behavior}
According to~\cite{Paul98ASR,Guo11EC}, a buyer's social network
strongly influences her purchasing behavior. In \cite{Guo11EC}, Guo
et. al. analyze data from the e-commerce site Taobao\footnote{Taobao
is a Chinese Consumer Market place, and also the world's largest
e-commerce website, http://www.taobao.com.} to understand how
individuals' commercial transactions are embedded in their social
graphs.  In the study, they show that implicit
information passing exists in the Taobao network, and that
communication between buyers drives purchases.
However, according to the study presented in
\cite{Jure07TWEB} social factors may impose a different level of
impact on the user purchase behavior for different e-commerce
products.

Several studies have been conducted to understand various aspects of
Groupon. In \cite{Arabshahi10Groupon}, Arahbshai examined the
business model of Groupon, and concluded that its advantages is the economic potential to leverage simple technologies (e.g., web portal and email subscription) to address deeply embedded inefficiencies in life. 
In \cite{Utpal10Groupon}, Utpal
conducted a survey-based study on Groupon, in order to understand
how businesses fare when running group promotions. Employee satisfaction, rather than features of the 
promotion or its effect, was found to be the factor that correlates most strongly with the 
profit gained from a promotion. Effectiveness in reaching new customers and the percentage of 
Groupon users who bought more than the deal's value during the visit were important 
factors for the small merchants when considering whether to run another promotion. 
In \cite{GrabchakBBG11WWW}, Grabchak et al. study the problem of selecting Groupon style chunked reward Ads. To address the problem, they devise several adaptive greedy algorithms in a stochastic Knapsack framework.  

The paper most related to our work is \cite{Byers11CoRR},
where data on the purchase history of Groupon deals
were analyzed. One key outcome of \cite{Byers11CoRR} is the
preliminary evidence that Groupon is behaving strategically to
optimize deal offerings, giving customers ``soft'' incentives (e.g.,
deal scheduling and duration, deal featuring, and limited inventory)
to make a purchase. Our work differs from these studies by focusing on
modeling the deal purchasing dynamics over time and by highlighting the importance
of the tipping point and its implication to social propagation.

\subsection{Collective Attention}

In \cite{Lerman10www,Lerman10icwsm,Lerman11tist}, Lerman et. al, propose to use a stochastic model to describe the social dynamics of web users, with Digg as a case study. The stochastic model focuses on describing the aggregated (by average quantities) behavior of the system, including average rate at which users contribute new stories and vote on existing stories. With the devised stochastic model, popularity of a Digg story can be predicted shortly after it was submitted (or with 10 to 20 votes). Studies in \cite{Lerman08sigcom,BakshyKA09EC,ColbaughG10ISI} have found that early diffusion of information within a community could be a good predictor of how far it will spread.   

Recent studies of collective attention on social media sites such as
Twitter, Digg and YouTube~\cite{FangB07PNAS,SzaboH10CACM,Ram11CoRR}
have clarified the interplay between popularity and novelty of user
generated content. The allocation of attention across items was
found to be universally log-normal, as a result of a multiplicative
process that can be explained by an information propagation mechanism inherent in all these sites. While the specific time scales over which novelty decays differ
between different systems depending on their typical type of content,
the functional form of the decay is consistent and thus future popularity is predictable.

\begin{table*}%[h]
\centering
\begin{tabular}{||l|c|c|c|c||}
\hline
Description  & coefficient  & standard error & t-value & p-value \\
\hline \hline 
Intercept & $-4.094\times10^{12}$ & $5.9776\times10^{12}$ & -0.6849 & 0.4935\\
Tipping Point & 0.7316 & 0.029 & 25.2276 & $6.5792\times{10^{ - 125}}$(***)\\
Featured position & 0.7004 & 0.0463 & 15.1189 & $2.0166\times{10^{-49}}$(***)\\
Duration & 0.0062 & $4.8862\times10^{-4}$ & 12.6412 & $1.6054\times10^{-35}$(***) \\
is limited or not & $-2.6105\times10^{-4}$ & $2.0969\times10^{-5}$ & -12.4494 & $1.5597\times10^{-34}$(***)\\
 Retail Price & - 0.0082& 0.0458 & -0.1797 & 0.8574 \\
Discount & -0.0011 & $1.6681\times10^{-4}$ & -6.3744  & $2.1908\times10^{-10}$(***)\\
\hline
\hline
Sunday & 0.0061 & 0.0022 & 2.7358  & 0.0063 (***)\\
\hline
\hline
Nightlife & 0.3208 & 0.1515 & 2.1180 & 0.0343 (*)\\
Health\&Fitness & 0.6429 & $0.0849$ & 7.5722  & $5.1827\times10^{-14}$(***) \\
Travel & -0.1789 & $0.0782$ & -2.2874  & 0.0223 (*)\\
Automotive & -0.3289 & 0.1366 & -2.4074  & 0.0161 (*)\\
Professional Services & 0.2552 & 0.1390 & 1.8363 & 0.0664 \\
\hline
\hline
atlanta & -2.0460 & 0.9373 & -2.1829 & 0.0291 (*)\\
albuquerque & -1.8548 & 0.9365 & -1.9806 & 0.0478 (*)\\
austin & -2.4329 & 0.9516 & -2.5567 & 0.0106 (*)\\
abbotsford & -2.1012 & 0.9392 & -2.2371 & 0.0254 (*)\\
barrie & -2.2454 & 0.9496 & -2.3646 & 0.0181 (*)\\
...&...&...&...&...\\
%anchorage & -2.2683 & 0.9528 & -2.3807 & 0.0174 \\
%akron-canton & -2.2447 & 0.9514 & -2.3595 & 0.0184 \\
%billings & -2.1431 & 0.9500 & -2.2559 & 0.0242 \\
%athens-ga & -3.0140 & 0.9625 & -3.1314 & 0.0018 \\
%bakersfield & -2.4164 & 0.9502 & -2.5431 & 0.0110 \\
%abilene & -2.8721 & 0.9526 & -3.0151& 0.0026 \\
%amarillo & -2.4938 & 0.9509 & -2.6226 & 0.0088 \\
%augusta & -2.8508 & 0.9793 & -2.911 & 0.0036 \\
%calgary & -2.4719 & 0.9483 & -2.6066 & 0.0092 \\
%birmingham & -2.1001 & 0.9369 & -2.2414 & 0.0251 \\
%corpus-christi & -2.1343 & 0.9470 & -2.2537 & 0.0243 \\
%buffalo & -1.8990 & 0.9513 & -1.9963 & 0.0460 \\
%columbia & -2.6489 & 0.9534 & -2.7783 & 0.0055 \\
%chattanooga & -2.1221 & 0.9584 & -2.2142 & 0.0269 \\
%fairfield-county & -2.3024 & 0.9580 & -2.4034 & 0.0163 \\
%denver & -2.6872 & 0.9477 & -2.8356 & 0.0046 \\
%el-paso & -2.0213 & 0.9448 & -2.1395 & 0.0325 \\
%fort-worth & -2.0186 & 0.9448 & -2.1366 & 0.0327 \\
%eugene & -2.3163 & 0.9599 & -2.4132 & 0.0159 \\
%fresno & -2.1139 & 0.9472 & -2.2318 & 0.0257 \\
%erie & -1.9444 & 0.9367 & -2.0758 & 0.0380 \\
%daytona-beach & -2.0563 & 0.9515 & -2.1611 & 0.0308 \\
%fort-myers-cape-coral & -2.0513 & 0.9466 & -2.1670 & 0.0303 \\
\hline
\end{tabular}
\caption{Multivariate linear regression of number of purchases. $N$ = 3876,
$R$-square = 0.5952, adjusted $R$-square = 0.5857. Note that, due to space limitation, we only show the result with p-value smaller than 5\% for the launching day, category and division study.}\label{tbl:correlation-attribute}
\end{table*}
% data section
\section{Datasets}\label{sec:data}
We collected data from Groupon's socially promoted and local daily deal websites in the US. 
We also collected data from LivingSocial to verify that our models could be applied more generally
across group deal sites.

Groupon provides a convenient API\footnote{http://www.groupon.com/pages/api},
which allows us to obtain more detailed information about the deals. 
By the end of April 2011, Groupon's business covered about 120 cities in the US\footnote{Statistics obtained from Groupon API.}. We monitored all Groupon deals offered in 60 different randomly selected cities during the period between April 4th and June 16th, 2011.
In total we collected the entire purchase traces of 4349 deals.

In LivingSocial, there is no API available for us to periodically obtain information about deals, 
so we developed a crawler to visit the webpages of deals periodically. After crawling for one month, 
we collected traces from over 900 deals. 

Next, to give a flavor of the type of data being used we examine the features of Groupon deals
in more detail. A similar examination for LivingSocial is outside the scope of this work.
However, we will later see that the models inferred from these observations apply to LivingSocial as well.

\subsection{Groupon Deal Characteristics}
%\begin{figure*}[thl]
%  \centering
%  \subfigure[Retail Price]{
%  \includegraphics[width=2.2in]{full_fig/bar-dist-retail-price.eps}
%  }
%  \subfigure[Discount]{
%  \includegraphics[width=2.2in]{full_fig/bar-dist-discount.eps}
%  }
%  \subfigure[Tipping Point]{
%  \includegraphics[width=2.2in]{full_fig/dist-tipping_point.eps}
%  }\\
% \subfigure[Tipping Time]{
%  \includegraphics[width=2.2in]{full_fig/dist-tipping_time.eps}
%  }
% \subfigure[Lifetime]{
%  \includegraphics[width=2.2in]{full_fig/dist-life.eps}
%  }
%  \subfigure[Final number of purchase]{
%  \includegraphics[width=2.2in]{full_fig/bar-dist-final-purchase.eps}
%  }
%  \caption{Distribution of different attributes of Groupon deals}\label{fig:dist-attributes}
%\end{figure*}
%\begin{table*}[htl]
%\centering
%\begin{tabular}{||l||c|c|c|c|c|c|c||}
%\hline
% Weekday & Sunday  & Monday & Tuesday & Wednesday & Thursday & Friday & Saturday \\
%\hline \hline 
%Proportion & 0.0724 &  0.1316 &   0.1555 &   0.1312 &   0.1642  &  0.2028 &   0.1423 \\
%\hline
%\end{tabular}
%\caption{Distribution of launching day}\label{tbl:launch_day}
%\end{table*}
At the time of our study the Groupon website presented the following relevant deal information: description, discount, 
time of launch, tipping point (purchases required for a deal to actually be sold), and the maximum number of sales of the deal. Additionally,
users could monitor the current number of purchases~\footnote{The current number of purchases has since our
study been removed and replaced with an obfuscated threshold to make it harder to make predictions.} and whether the deal has tipped or sold out.
%Each deal is associated with a basic set of features: the deal
%description, the retail and discounted prices, the start and end
%dates, the ``tipping point" required for the deal to be ``on", the
%number of coupons sold, whether the deal was available in limited
%quantities, and if it sold out.
We monitored the number of purchases and the position of each deal
in 20-minute time intervals.
A surprisingly large portion (10\%) of all deals exhibited dramatic non-monotonically increasing
behavior, e.g. a decrease of 10 purchases between subsequent intervals. This may indicate that something was wrong with the deal, e.g. false marketing due
to an inflated list price, and customers who initially purchased the deal requested a refund (an option
Groupon supports and markets). Due to the unknown user behavior behind these deal actions we exclude these
deals from our study. %We, however, intend to look more closely at this phenomena in future work.
Hence, 3876 deals were left to analyze.
%\begin{table}[htl]
%\centering
%\begin{tabular}{|l||l|}
%\hline
%Category & Proportion \\\hline
%Arts\&Entertainment &  0.1667 \\
%Beauty\&Spas &  0.1050 \\
%Food\&Drink &  0.0849 \\
%Nightlife &  0.0135 \\
%Pets & 0.0050 \\
%Shopping&   0.1517\\ 
%Home Services& 0.0862 \\  
%Education &  0.0172 \\
%Restaurants &  0.1695 \\
%Health\&Fitness & 0.1047 \\
%Travel &  0.0175 \\
%Automotive & 0.0188\\
%Professional Services &  0.0592\\
%\hline 
%\end{tabular}
%\caption{Distribution of categories}\label{tbl:categories}
%\end{table}
In our dataset, 270 deals (out of 3876) had not reached their tipping point when they expired.
In the following, these deals are called \emph{failed}
deals; and deals that are turned on successfully are called
\emph{tipped} deals.

\subsubsection{Attributes of Deals}
Here we present some statistics about attributes of the deals in our Groupon dataset, 
including retail price, discount, deals needed to tip (tipping point), time
needed to tip (tipping time), lifetime of a deal
and final number of purchases.

Groupon deals have different retail prices and discounts.
%, and the
%statistics for our collected dataset are shown in
%Figure~\ref{fig:dist-attributes} (a) and (b). 
The mean value of
 retail price is \$44 and the mode value is \$10. We
observe that most of the discounts range from 50\% to 60\%, and the mode
value is 50\%. Based on these statistics, we see that the product and
services deals provided on the Groupon website are not expensive most
of the time, and the discounts are usually very big.

In Groupon, deals may have different tipping points and successful
deals may also have different tipping times even when they have the
same tipping points. 
%Distributions of tipping points and tipping
%times over tipped deals are shown in
%Figure~\ref{fig:dist-attributes} (c) and (d). 
The average number of tipping points or units needed to tip is 22 (mode value is around 10)
and the expected tipping time is about 10.5 hours (mode
value is around 6.67 hours). Most of the time, deals in Groupon were
tipped within one day.

%We plot the distribution of lifetimes of deals and the
%distribution of the final number of purchases of deals in
%Figure~\ref{fig:dist-attributes} (e) and (f), respectively.  
Note
that the lifetime of a deal in Groupon is usually set to 1 day,
2 days, 3 days or 4 days. The average number of purchases of a
deal is 373.  A deal may be specified with a limited available
quantity. So these numbers are mixtures of different
factors, such as the quality of a deal itself, the quantity available etc.

%We show the launching day (i.e., the day when the deal is posted on the Groupon website) of a deal in Table~\ref{tbl:launch_day}, and find that Groupon schedules many deals on Thursday and Friday. A possible reason is that people usually begin to plan their weekend activities around Thursday and Friday, so the timing is good for promoting deals. Note that the percentage of deals launched on Sunday is relative small. %The explanation is that, people begin to work on Monday and probability are not interesting in planning any activities. Thus intensity for promoting deals is reduced.  

%Finally, we show the categories of deals promoted in Groupon in Table~\ref{tbl:categories}. Most of the deals are related to art and entertainment,  beauty and spas, shopping, restaurants and health, and fitness. 

%\begin{figure*}[htb]
%  \centering
%  \subfigure[Launching at around 4:00am]{
%    \includegraphics[width=2.1in]{full_fig/groupon-tipping-time-startingtimes-4.eps}}
%  \subfigure[Launching at around 5:00am]{
%    \includegraphics[width=2.1in]{full_fig/groupon-tipping-time-startingtimes-5.eps}}
%  \subfigure[Launching at around 6:00am]{
%    \includegraphics[width=2.1in]{full_fig/groupon-tipping-time-startingtimes-6.eps}}
% \caption{Tipping time distributions of deals with different launching times in Groupon}\label{fig:start}
%\end{figure*}

\subsubsection{Factors Impacting Purchases}
As we are ultimately interested in modelling purchase dynamics of deals, 
we first need to understand what factors impact purchases.
Hence, we regress the attributes discussed in the previous section against the final number of purchases of a deal.
If the Groupon commission is known\footnote{reportedly 50\% in \cite{Arabshahi10Groupon}}, this number 
also gives a good estimate of the merchant's revenue from a deal.
%The regression model can facilitate us to quantify the correlation of the final number
%of purchases and the various deal attributes, thus further help us find the important attributes which are critical to understand the purchase dynamics of group deals.

The model we use is as follows. Let $N_L$
denote the final number of purchases, $\theta$ the number of purchases needed to tip (tipping point), $f$ whether the deal is listed in featured position (1) or not (0) at the current time, $L$ the time till the $N_L$-th purchase, $p$ the retail price, $d$ the discount, and finally $l$ whether the deal inventory is limited (1) or not (0). The parameters $\textbf{w}$, $\textbf{c}$ and $\textbf{g}$ are vectors encoded as in~\cite{Byers11CoRR} to represent the launch day, category, and city. The following equation is also taken
from~\cite{Byers11CoRR}.
%Dummy-coding refers to using binary vectors to encode categorical variables, where a variable that can take on $k$ distinct values is encoded using a binary vector of length $k-1$ where at most one entry is set to one. 

\begin{equation}
\begin{split}
 \log N_L  & = \beta_0 + \beta_1  \log \theta + \beta_2 f+\beta_3 L + \beta_4 l + \beta_5 p + \beta_6 d \\ & + \overline{\beta}_7 \textbf{w} + \overline{\beta}_8 \textbf{c} +\overline{\beta}_9 \textbf{g}
\end{split}
\end{equation}
where $\beta_0 \sim \beta_9$ are the coefficients of the linear model.

We fitted the model using multivariate linear regression. The parameter estimates, their standard errors, t-values and p-values are listed in Table~\ref{tbl:correlation-attribute}.  Due to space limitations, only attributes with significance level (p-value) smaller than 5\% are shown in the table.
Among those attributes, we find that tipping point and featured position are the two most significant factors that can help predict the number of purchases. Surprisingly, tipping point seems to have better predicting power than featured position (i.e., the t-value is much larger for the tipping point factor than for the featured position factor). 
In the next section, we show how the tipping time can be generalized 
as an inflection point in the purchase dynamics of group deals.

% dynamics

\section{Purchase Dynamics}\label{sec:model}
In this section, we propose a model of the purchase
dynamics of group deals.
A group deal
is generally discovered by the user in one of the following four
ways: (1) by visiting a web-page, (2) by running a
smart-phone application, (3) by getting notifications via email and
(4) by communicating with friends. Here, we refer to the first three as random discovery and the fourth is referred to as social propagation.

Based on this notion, our model describes the purchase dynamics as follows. Let $N_t$ denote the number of times that the deal has been purchased at time $t$. We then have
\begin{equation}\label{eqn:general}
N_{t + \Delta t} - N_{t} = \alpha_t \cdot Y_t + \beta_t \cdot f(t,N_t),
\end{equation}
where $\alpha_t$ and $\beta_t$ are weight factors, $Y_t$ is a non-negative random variable denoting the number of purchases caused by random
discovery in the interval $(t, t + \Delta t]$, and $f(t,N_t)$ represents the number of purchases caused by social propagation in the same interval as a function of $t$ and $N_t$.

%Note that there is no tipping point in LivingSocial, the incentive that drives users to buy deals is the following commitment made by LivingSocial: ``Buy first, then share a special link with friends, if three friends buy, yours is free!".  So it will be of great interests to see how the purchase dynamics changes in LivingSocial against Groupon. 

\begin{figure}[htb]
  \centering
\subfigure[Groupon]{\includegraphics[width=1.6in]{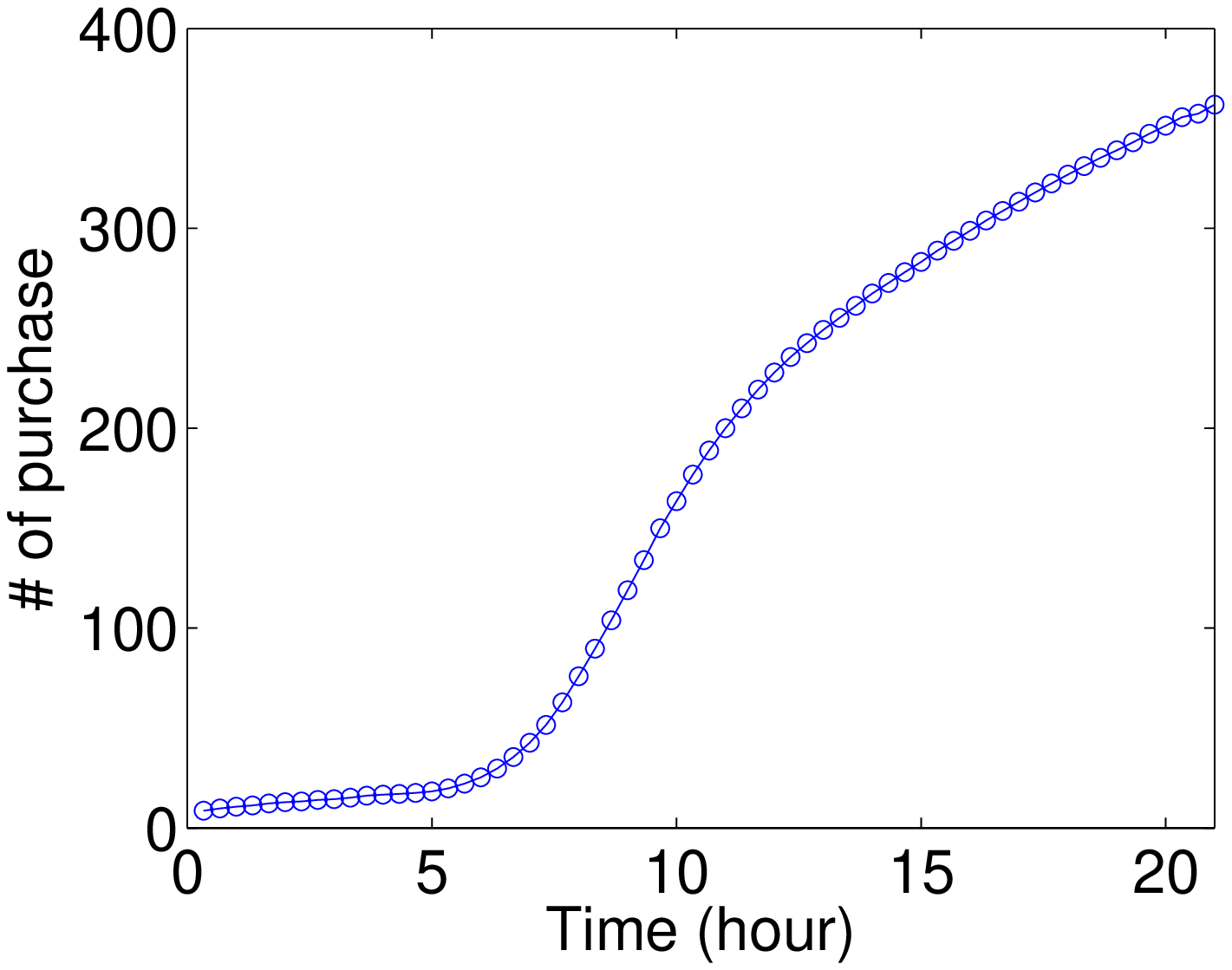}}\hspace{-10pt}
 \subfigure[LivingSocial] {\includegraphics[width=1.6in]{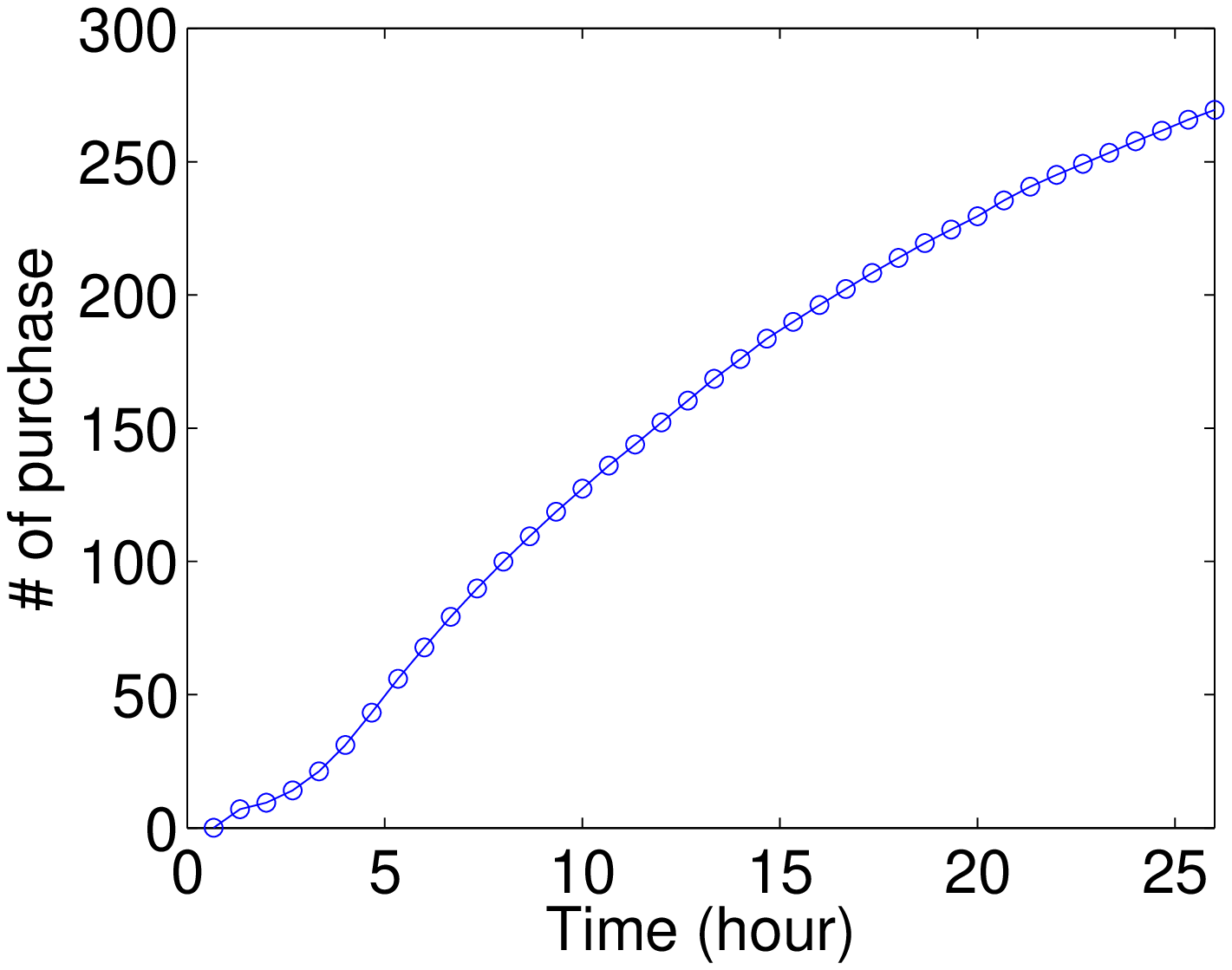}}   
\caption{Purchase growth of deals}\label{fig:purchase_growth} 
\end{figure}

\begin{figure}[htb]
  \centering
\subfigure[Groupon]{\includegraphics[width=1.6in]{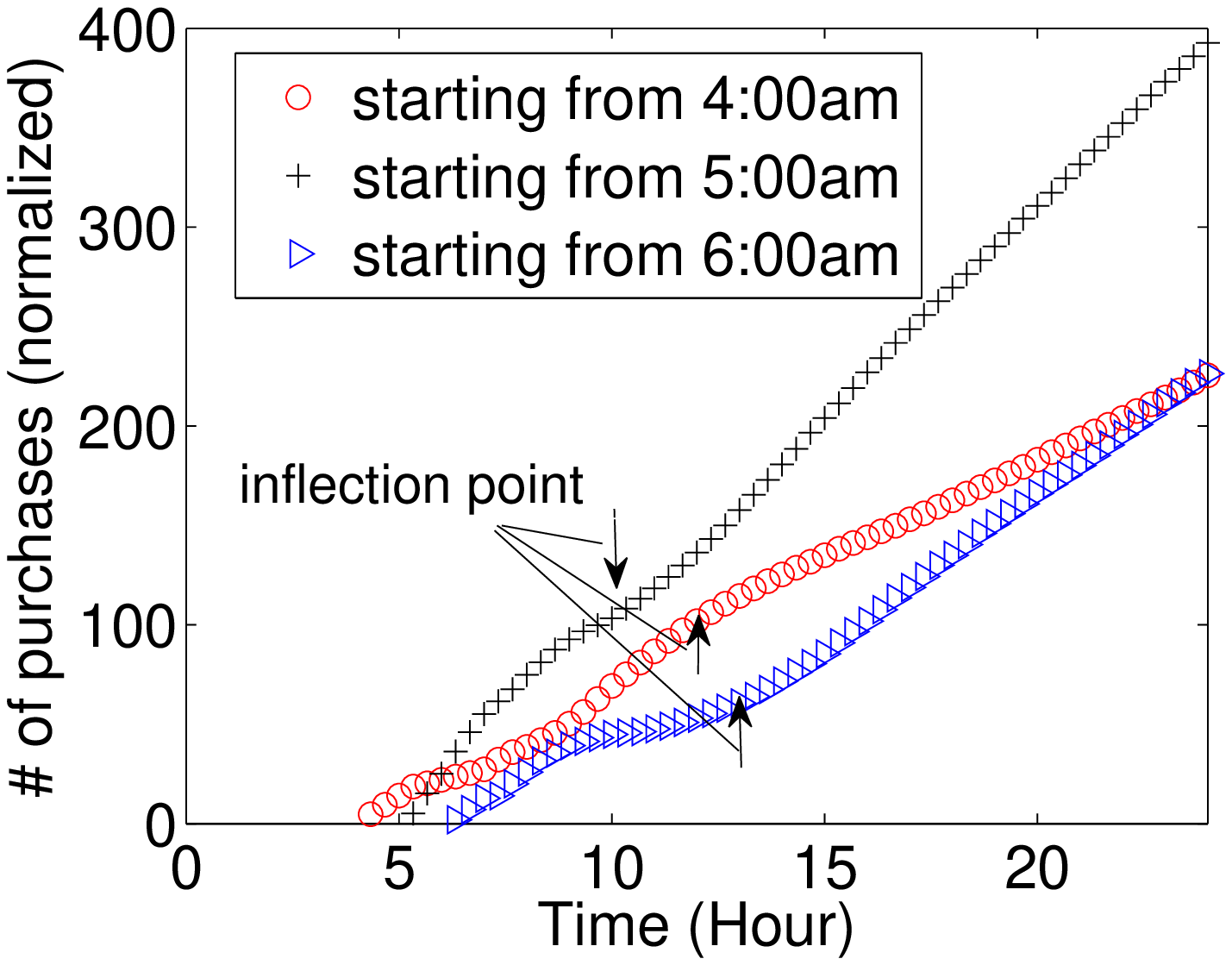}}\hspace{-10pt}
 \subfigure[LivingSocial] {\includegraphics[width=1.6in]{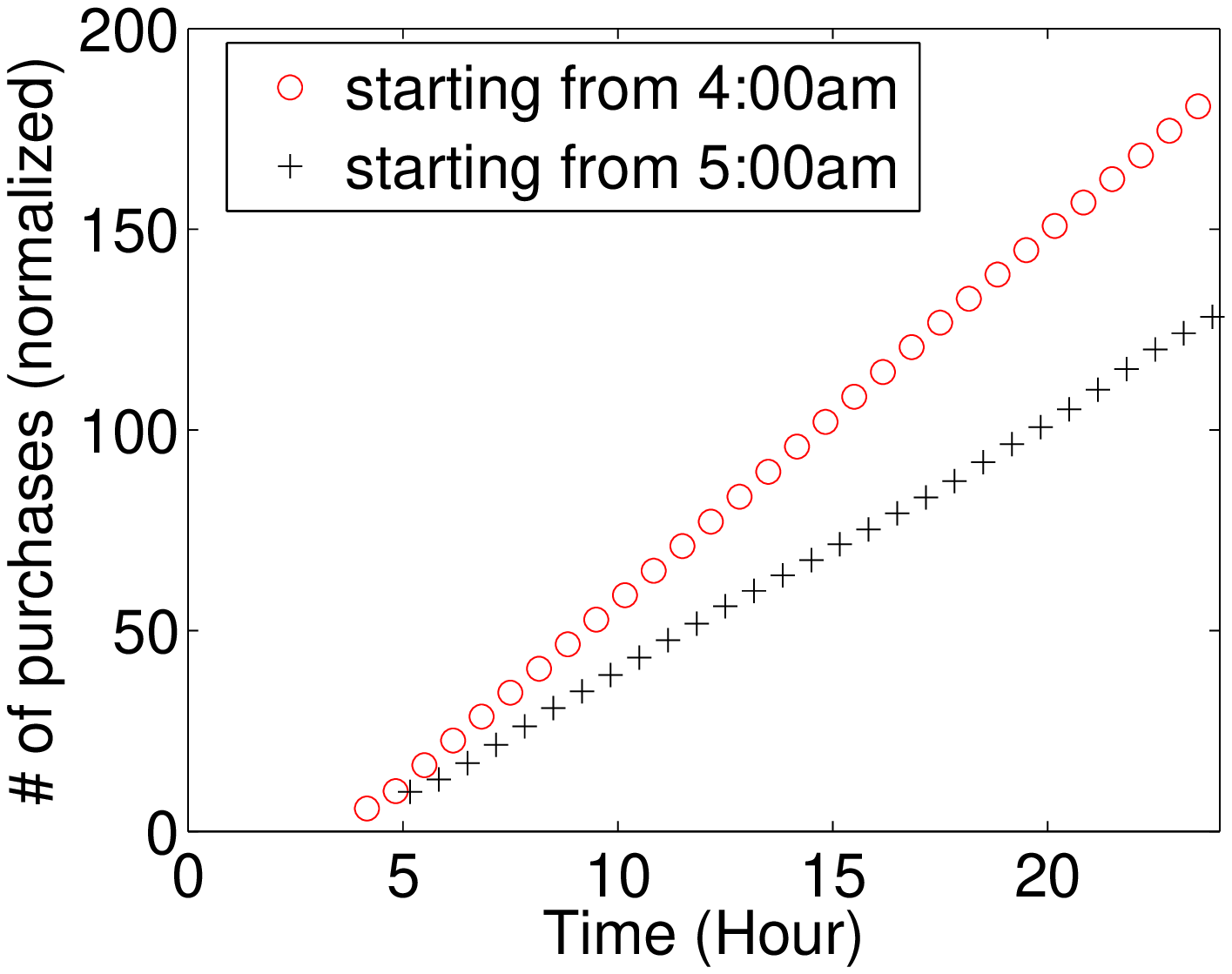}}   
\caption{Normalized Purchase growth of deals}\label{fig:purchase_growth_normalize} 
\end{figure}

We average the number of purchases of deals for each time step in both Groupon and LivingSocial. As shown in Figure~(\ref{fig:purchase_growth}), deals in LivingSocial grow faster than Groupon in the first few hours. A possible reason is due to the different incentive that LivingSocial is using to promote deals. LivingSocial users who want to get free deals may disseminate deal information more eagerly.   

Furthermore, there is an inflection point in the purchase dynamics for both Groupon and LivingSocial deals (after around 7 and 4 hours in Figure~\ref{fig:purchase_growth}(a) and (b), respectively), after which the number of purchases grows faster; whereas  the number of purchases grows relatively slowly and steadily before the inflection point. Note that after the inflection point, the number of purchases grows dramatically for about 11.6 and 14.8 hours in Groupon and LivingSocial, respectively, after which the purchase rate drops.
%Based on the observation, we find the purchase growth in LivingSocial exhibits a similar behavior as in Groupon.%,  where multiplicative process can be utilized to capture the dynamics. %We consider the purchase dynamics after 4 hours in LivingSocial follows multiplicative process. 
%So similar to the study of after tipping in Groupon, we study the purchase dynamics (after around 4 hours) of LivingSocial as follows. 

One may argue that this inflection point could be caused by time-of-day seasonality given that all deals 
are local for a region belonging to a single time zone. For example, most people
do not buy deals at night, but early in the morning when they wake up. Hence, we normalize the number of purchases by removing the seasonal impact to examine whether the inflection 
point is caused by the time the deal is launched, as shown in Figure~\ref{fig:purchase_growth_normalize}.  In Groupon, 95\% of the deals are launched before 7:00am and 50\% of these are launched between 4:00am and 6:00am.
Hence, we cluster deals in three groups, those that launch around 4:00am, 5:00am, and 6:00am respectively. As shown in Figure~\ref{fig:purchase_growth_normalize}(a), normalized purchase growth of deals clearly has two-stage growth, which is divided by a inflection point. Before the inflection point, it shows non-linear growth; while after the inflection point, it obeys linear growth. In LivingSoical, deals are launched during 4:00am$\sim$6:00am, like Groupon. Interestingly, in Figure~\ref{fig:purchase_growth_normalize}(b), we find the inflection point in the purchase growth of LivingSocial deals disappears after the normalization. In addition, deals launched from the same time (e.g., from 4:00 am) exhibit different purchase dynamics behavior in Groupon and LivingSocial, e.g., in Figure~\ref{fig:purchase_growth_normalize}, the purchase dynamics of Groupon deals still exhibit an inflection point, while there is none in LivingSocial deals.  
These observations suggest that: (1) the consistent launch times may cause the two-stage purchase growth in LivingSocial; but (2) the inflection 
point cannot solely be attributed to the time the deal is launched in Groupon, but the tipping-point mechanism may also play a role here.  

%If the deals are launched at different times and 
%reach their inflection points at different times, but the time it takes to reach the inflection points remains
%constant, then we argue that the inflection point is not caused by time-of-day seasonality but it is rather an artifact of the deal gaining momentum after some time in the market. On the other hand if the inflection points are reached
%at the same time-of-day irregardless of the time the deals were launched we would argue that the inflection
%point is caused by time-of-day seasonality.
%
%In Groupon, 95\% of the deals are launched before 7:00am and 50\% of these are launched between 4:00am and 6:00am.
%Hence we cluster deals in three groups, those that launch around 4:00am, 5:00am, and 6:00am respectively.
%We then examine if their average tipping times are the same or if they are adjusted to the time-of-day cycle.
%Figure~\ref{fig:start} shows that the former is the case, and hence the inflection point is not caused
%by time-of-day seasonality. The mode value of tipping time for the three groups are $6.67$,$6.33$ and $6.67$
%hours. 

Based on the above observations we write our equation as:
\begin{equation}\label{eqn:groupon}
N_{t + \Delta t} - N_t =
\begin{cases}
Y_t  & \mbox{before the inflection point}\\
r(t) X_t N_t  & \mbox{after the inflection point}
\end{cases}
\end{equation}
Thus, we are implicitly assuming that before the inflection point $\alpha_t = 1$ and $\beta_t = 0$, whereas after the inflection point $\alpha_t = 0$ and $\beta_t = 1$ in \eqref{eqn:general}.
This assumption is motivated by the fact that random discovery dominates before the inflection point and social propagation dominates afterwards --- even though the two processes may coexist. In particular, before the inflection point the customer base is
small so the random discovery process dominates. In addition, in Groupon, before the deal has tipped, people will hesitate to make a purchase, as
it is still uncertain both whether the deal was considered good by others and whether it will be offered, which reduces the effects of social propagation. 
After the inflection point both of these uncertainties are gone.

According to \eqref{eqn:groupon}, after the inflection point, the increase in the number of purchases ($N_{t + \Delta t} - N_t$) is proportional to the number of people that has purchased the deal up to time $t$.  Intuitively, a fraction of the people that already purchased the deal will notify some of their friends about it, and a fraction of these friends will purchase the deal.  These fractions are represented by the positive random variable $X_t$.  We assume that $\{X_t\}$ are independent and identically distributed random variables.  Since $X_t$ is assumed to be positive, $N_t$ can only increase over time.  This growth in time is eventually curtailed by a decay in novelty, which is parameterized by the factor $r(t)$.
As we discuss later, $r(t)$ is decreasing in $t$. This notation of social 
propagation is borrowed from and motivated in more depth in~\cite{FangB07PNAS}.

\subsection{Purchase Dynamics Before Inflection}
We denote by $\tau_i$ the interarrival times of purchases.  In particular, $\tau_i$ is the time between the ${i-1}$ and the $i$-th purchases.
Suppose that each $\tau_i$ is independently drawn from some distribution $F$.  We denote a deal's inflection point by $\theta$, that is, the number of purchases required before social propagation dominates.  Let $L$ be the total time that the deal is open for purchases (as set by the
seller).
Then, $N_L$ is the final number of purchases when the deal ends.

Let $F_n$ denote the $n$-fold convolution of $F$.  Then, $F_n$ is the distribution of the sum of $n$ consecutive interarrival times.  Thus, the distribution of the time span to get the same inflection point $\theta$ for deals is given by  $F_{\theta}$, the $\theta$-fold convolution.

\begin{figure}[htl]
\centering
\includegraphics[width=2.9in, height = 2in]{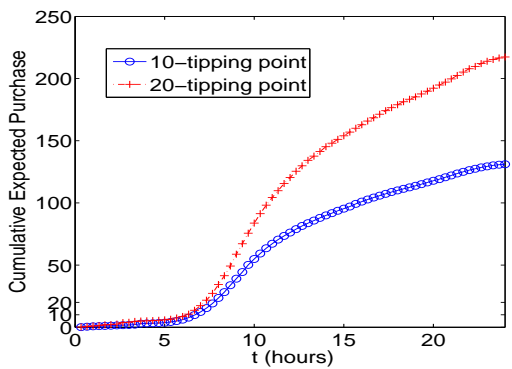}
\caption{Purchase growth for deals with tipping (inflection) points of 10 and 20, respectively, in Groupon}\label{fig:purchase-growth}
\end{figure}

Figure~\ref{fig:purchase-growth} shows how the number of Groupon deal purchases increases over time when the tipping point is equal to 10 (the most frequent value) and 20 purchases. The plot is based on 492 (resp. 477) deals whose tipping point was equal to 10 (resp. 20) in our dataset.   We observe the same pattern for deals with other tipping points, e.g., 5 and 30.
We find an approximately linear growth of purchases at the beginning of the lifetime of a deal.  For both tipping points, the purchase rate
is relatively small and steady before the tipping time.  After tipping or around tipping, the number of purchases grows dramatically for about 11.6 hours, after which the purchase rate drops.  The tipping point time, thus, typically coincides with the inflection point time in the purchase dynamics.

Note that the final number of purchases of a deal with a tipping point of 10 purchases is usually smaller than the corresponding number for a deal with a tipping point of 20, even though we do not observe a significant difference before the tipping times. One possible reason is that deals tipping after 10 purchases have smaller purchase populations than those that tip after 20 purchases, depending on the specific categories of products and services. Furthermore, the potential purchase population may also act as the reference for Groupon and local merchants when they set the tipping point for a deal.

We now look at the probability that a deal fails, i.e. does not reach the inflection point.
We say that a deal is turned
on as long as its number of purchases reaches the inflection point
$\theta$ before the deal expires, i.e. its lifetime $L$ ends. So the
probability of a deal failing is equal to $\Pr(N_L<\theta)$.

\begin{equation}\label{eqn:fail-prob}
\Pr(N_L<\theta)=\sum\limits_{n = 1}^{{\theta} - 1} {\Pr (N_L = n)} \
\end{equation}

Since the $\tau_i$ variables are iid interarrival times of purchases,
it follows that this is a renewal process. We use ${S_{n}} =
\sum\limits_{i = 1}^{n} \tau_i$ to denote the time spent until
the $n{th}$ purchase.

It is easy to see that $N_t = \sup\{n: S_n \le t \}$, and thus,
\begin{equation}\label{eqn:renewal}
\begin{split}
&\Pr(N_t=n) \\= &\Pr(N_t \ge n)- \Pr(N_t \ge n+1)\\= &\Pr(S_n \le
t)-\Pr(S_{n+1} \le t)\\=& F_n(t)-F_{n+1}(t)
\end{split}
\end{equation}

Applying Equation~(\ref{eqn:renewal}) to
Equation~(\ref{eqn:fail-prob}), we have:
\begin{equation}\label{eqn:tipping}
\begin{split}
&\Pr(N_L<\theta) \\
=&\sum\limits_{n = 1}^{{\theta} -
1}{(F_{n}(L)-F_{n+1}(L))}\\
= & F(L)-F_{\theta}(L)
\end{split}
\end{equation}

Note that Equation~(\ref{eqn:tipping}) can predict the
failure ratio (i.e., the probability not to be turned on) of a deal.
Conversely, using this equation, given the failure ratio, we can
estimate the parameters of $F$, such as the mean value.

\begin{figure}[htl]
\centering
\includegraphics[width=3in]{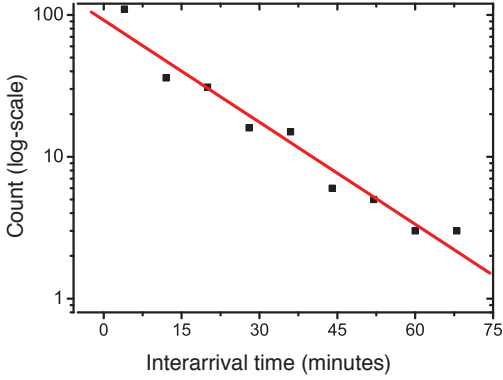}
\caption{Distribution of waiting time for a purchase. This
result is based on all deals with a tipping
point of 10 purchases, in Groupon}\label{fig:timeinterval}
\end{figure}

This analytical model can be easily extended to predict the
probability that a deal will be turned on when we know the number of purchases up to a given point in time. For
example, if at time $t_1$, a deal has already got $n_1$ purchases,
then the probability that the deal will be turned on can be
estimated as
\begin{equation}
\begin{split}
\Pr(N_L < \theta | N_{t_1} = n_1) = F(L-t_1) - F_{\theta-n_1}(L-t_1)
\end{split}
\end{equation}

We now consider what distribution the interarrival times follow in Groupon.
To exclude the impact of tipping point differences, we first consider only deals with a tipping point of 10 purchases (the tipping
point distribution mode) from all the data we gathered. As shown in
Figure~\ref{fig:timeinterval}, interarrival times follow an exponential
distribution. Thus, before tipping, the arrival rate of purchases follows a Poisson process.

This observation confirms our assumption about random discovery,
since if a user randomly checks the websites or a smartphone app the
probability of a purchase taking place in the next infinitely-small
time interval is the same, and hence the intervals between purchases follow an exponential distribution.  The Exponential fit in Figure~\ref{fig:timeinterval} has $R^2$ value 0.9784. We also check the interarrival times of purchases in LivingSocial during the first 4 hours, and find that interarrival times in LivingSocial also follow an exponential distribution.
%gives an average duration of 36
%minutes of $x_i$.

\begin{figure}[htl]
\centering
\includegraphics[width=3.3in]{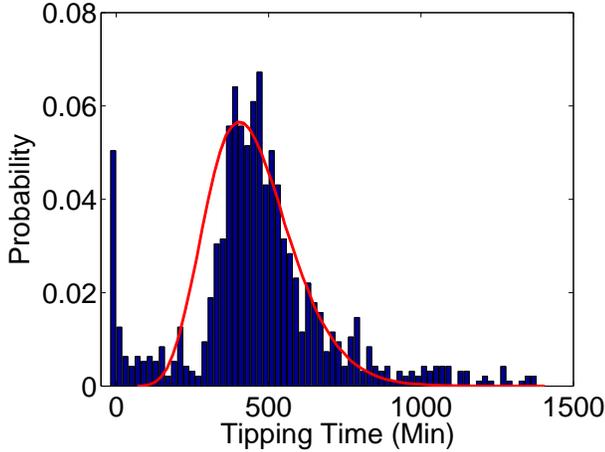}
\caption{Predicted tipping time distribution vs. empirical tipping
time distribution. The result is based on all deals with tipping
point equal to 10, in Groupon}\label{fig:tippingtime}
\end{figure}

%\begin{figure}[htl]
%  \centering
%  \subfigure[Density plot] {\includegraphics[width=1.7in]{full_fig/Empirical_Simulate_Compare.eps}} \hspace{-15pt}
%  \subfigure[QQ-plot]{
%    \includegraphics[width=1.7in]{full_fig/QQPlot.eps}}
% \subfigure[QQ-plot for the data, of which appealing deals are removed]{
%    \includegraphics[width=1.7in]{full_fig/QQPlot2.eps}}
% \caption{Predicted tipping time distribution vs. empirical tipping
%time distribution. The result is based on all deals with 10 tipping
%point.}\label{fig:tippingtime}
%\end{figure*}

%With the distribution of $x_i$, we could predict the failure
%probability according to the results from the previous section, or we
%could compare the expected value of inter-arrival arrival computed from failure
%probability with the mean of $x_i$. From our data, the expected
%value of inter-arrival time computed from failure probability is around 52
%minutes. The difference from the actual fitted value may come from
%the impact of other factors not considered in the model.

An important conclusion from our model is that the distribution of
tipping time in Groupon is expected to follow an n-fold convolution of distributions of $F(t)$.
Now, given that $F(t)$ is following an exponential distribution,
then deals with a tipping point of 10 purchases should follow a Gamma distribution
with a shape factor equal to 10. We compare the predicted
distribution of tipping time with that of real values gathered
online, the histogram and PDF (curved line) of the empirical and modelled distributions respectively are shown in
Figure~\ref{fig:tippingtime}. Note that there are some deals in Groupon that are very appealing, and thus were tipped immediately after they were launched. Nevertheless, the predicted tipping time distribution of Groupon deals is similar to the empirical one. %It could be seen that the theoretical
%prediction agrees with real data. Note that there is discrepancy in the prediction for the tipping time between 0 minutes and 200 minutes. After examining the density plot of empirical data in Figure~\ref{fig:tippingtime}(a), there are some deals are very appealing thus  were tipped immediately after they were launched.  After removing those appealing deals, i.e. deals with less than 60 minutes tipping points are removed, we get another QQPlots as shown in Figure~\ref{fig:tippingtime}(c). In practice, there should be some deals tipping very fast, but not that many. So to clear all the appealing deals is also problematic. Furthermore, let's think about whether delayed renewal process can help. The general idea about delayed renewal process is to extend the inter-arrival time for the first arriving buyer. So If apply delayed renewal process, we can not alleviate the discrepancy shown in the figures.

\subsection{Purchase Dynamics After Inflection}
We now focus on the dynamics after the inflection point, and for expositional clarity consider the time of inflection as time 0. 
Thus, $N_0$ denotes the number of purchases of a deal at the inflection point time.
Then, according to Equation~(\ref{eqn:groupon}), the number of purchases at time $T$ (that is, $T$ time units after the inflection point) is given by

\begin{equation}\label{eqn:dynamics}
N_T = \prod_{t=1}^T(1+r(t) X_t)N_0
\end{equation}
Note that the realization of $X_t$ will in general be different in different time periods; however all random variables $X_t$ follow the same distribution. When $X_t$ is small (which is the case for small time steps),
we have the following approximate solution for $N_T$:
\begin{equation}\label{eqn:approximate}
N_T \approx \prod_{t=1}^T e^{r(t) X_t}N_0 = e^{\sum_{t=1}^Tr(t) X_t} N_0.
\end{equation}
Taking the logarithm on both sides, we get
\begin{equation}\label{eqn:log_approximate}
\log N_T - \log N_0 \approx \sum_{t=1}^T r(t) X_t
\end{equation}

\begin{figure}[htl]
  \centering
  \subfigure[Groupon]{
  \includegraphics[width=1.65in]{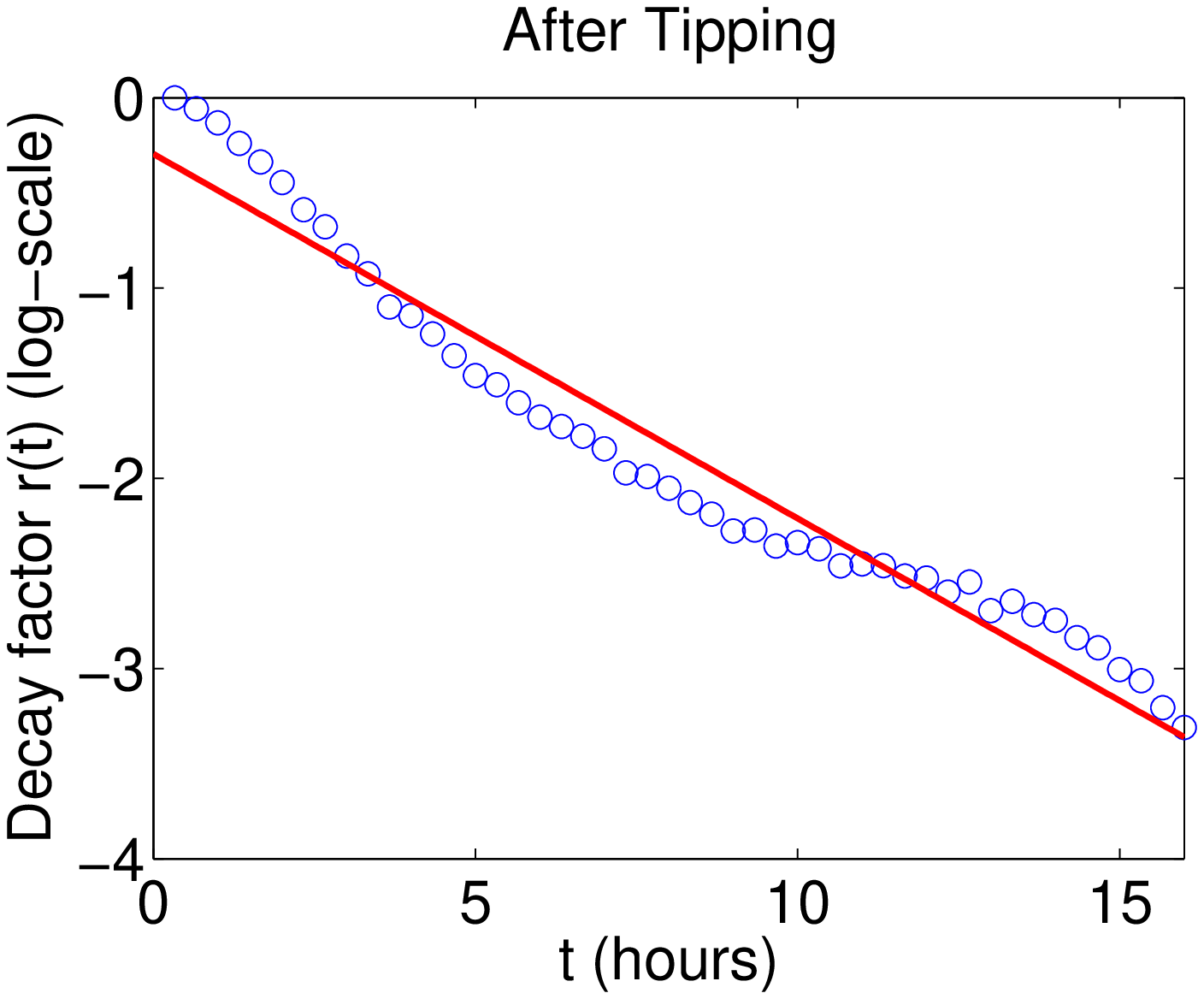}
  }\hspace{-20pt}
  \subfigure[LivingSocial]{
  \includegraphics[width=1.65in]{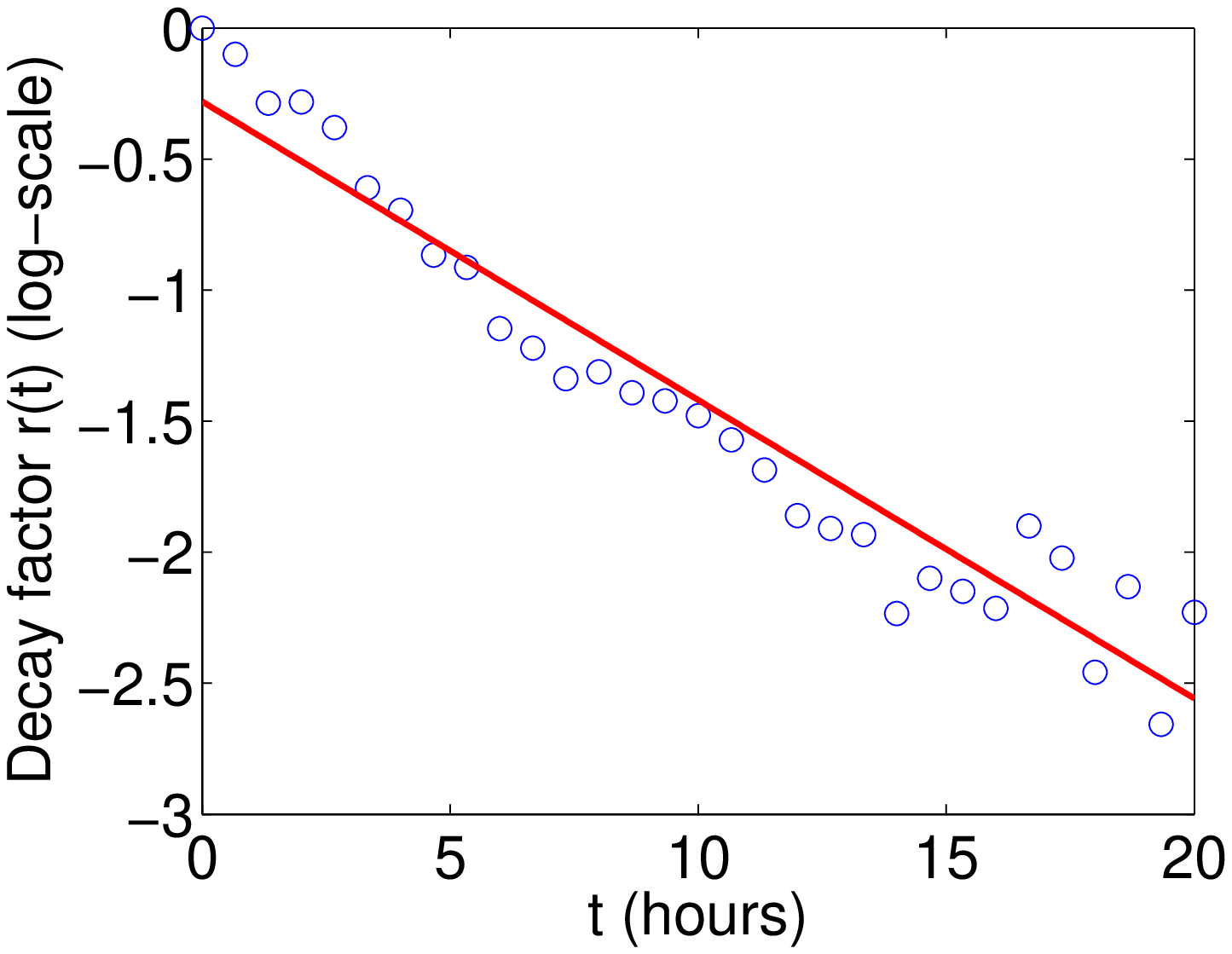}
} 
  \caption{Process of novelty decay}\label{fig:novelty}
\end{figure}

%\begin{figure}[htl]
%\centering
%\includegraphics[width=3.5in]{fig/prediction-growth.eps}
%\caption{Empirical verification of our model}\label{fig:growth-prediction}
%\end{figure}

The decay factor $r(t)$ is estimated according to Equation (\ref{eqn:groupon}) and Equation (\ref{eqn:log_approximate}) as follows:
\begin{equation}
r(t) = \frac{\E(\log N_t) - \E(\log N_{t-1})}{\E(\log N_1) - \E(\log N_0)}
\end{equation}
where we normalize $r(1)$ to 1. This calculation is again borrowed from and evaluated in
more detail in~\cite{FangB07PNAS}.

\begin{figure}[htl]
  \centering
  \subfigure[Groupon]{
  \includegraphics[width=1.5in]{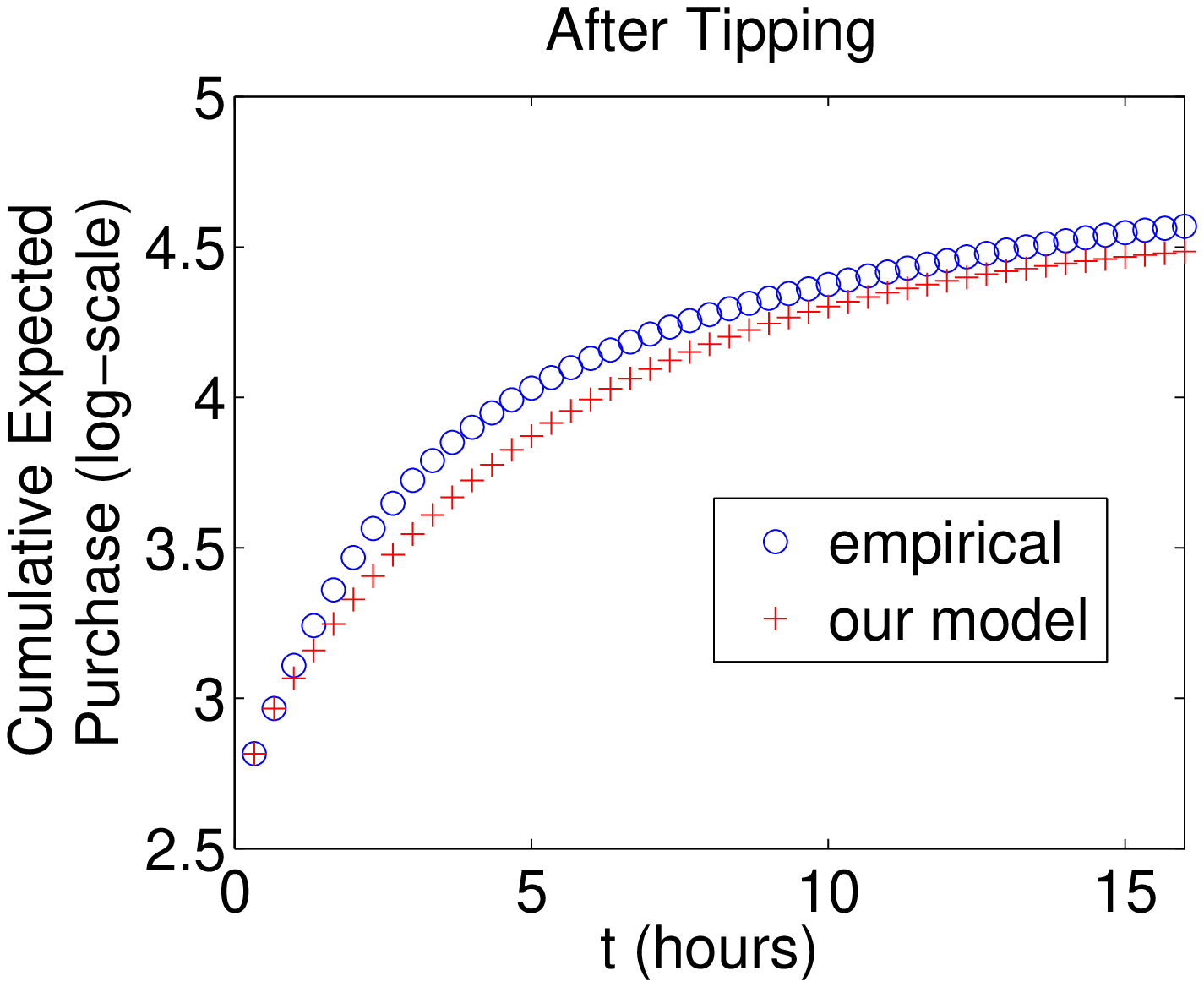}
} \hspace{-5pt}
  \subfigure[LivingSocial]{
  \includegraphics[width=1.5in]{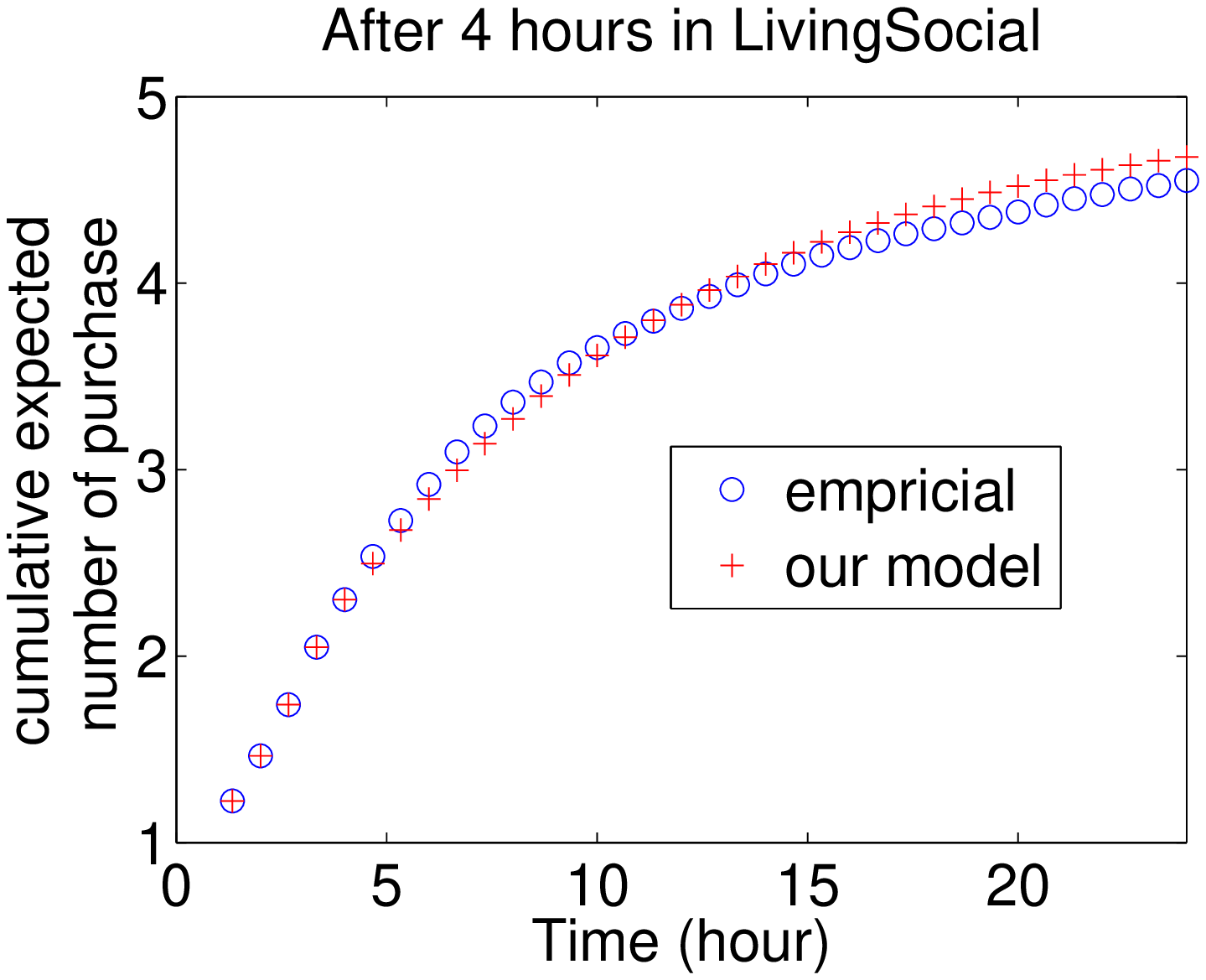}
  }
  \caption{Empirical verification of our model}\label{fig:growth-prediction}
\end{figure}

In Figure~\ref{fig:novelty}, we plot the novelty decay $r(t)$ for the first 16 and 20 hours after the inflection point in Groupon and LivingSocial, respectively, as estimated from our dataset. Note that tipping time is usually around 8 hours, so we focus on the time duration of 16 hours after tipping in Groupon. Recall that in this section $N_0$ denotes the tipping
point, and time $t = 0$ is the tipping time.  We observe that $r(t)$ decreases over time.  
Moreover, Figure~\ref{fig:novelty}, suggests that the novelty decay is exponential.  In particular,
\begin{equation}
r(t) \approx \exp(at+b),
\end{equation}
where in Groupon $a = -0.21$ and $b=-2$, and the $R^2$ value for this fit is 0.8839; and in LivingSocial $a = -0.11$ and $b = -0.28$ and $R^2$ value for this fit is 0.9190.

Next, we are interested in evaluating how well our model helps
explain the purchase growth after a deal has turned on.
With both $a$, $b$ estimated, we can use our
results to explain the growth of purchases.  In
Figure~\ref{fig:growth-prediction}, we demonstrate the potential predictive power of our model by empirically verifying the growth of purchases of deals after they have tipped. For the model fitting in Figure~\ref{fig:growth-prediction}, the $R^2$ value is 0.9404 and 0.9903 in Groupon and LivingSocial, respectively.

%\subsection{Purchase dynamics in LivingSocial}
%
%According to Equation~(\ref{eqn:log_approximate}), we calculate the decay factor for different time steps (starting from the 4th hour), and plot the result in Figure~\ref{fig:novelty-LS}(a). Note that in Figure~\ref{fig:novelty-LS}(a), decay factor values are in log-scale. Similar to Groupon deals, the decay factor in LivingSocial deals also follows an exponential decay, i.e., exp(ax+b) based on the curve fitting over the empirical data, where in Figure~\ref{fig:novelty-LS}(a), 
%
%
%
%
%Furthermore,  we are interested in evaluating how well our model helps
%explain the purchase growth of a deal after around 4 hours.
%With estimated values of $a$ and $b$, (i.e., $a = -0.074$ and $b = -0.27$ in Figure~\ref{fig:novelty-LS}(a)), we can use our results to explain the growth of purchases. 
%In Figure~\ref{fig:novelty-LS}(b), we demonstrate the potential predictive power of our model by empirically verifying the growth of purchases of deals after around 4 hours.  $R^2$ value for the fit in Figure~\ref{fig:novelty-LS}(b) is 0.9903. 

% prediction section
\section{Purchase Prediction}\label{sec:prediction}
In this section, we discuss how to use our models to predict the number of purchases of deals at a given time. Purchase prediction is important for both group deal websites and local merchants. Accurate forecasts may help group deal websites design more optimized deal scheduling and promotion strategies and aid local merchants in allocating
resources  more efficiently.

%There are two desired criteria for a devised prediction method: 1) the method should be able to perform prediction as early as possible 2) the method should be able to perform prediction as accurate as possible. Therefore, 
We now discuss methods which make predictions based on $h$ hours of previous observations.

\subsection{Predictors}

\subsubsection{Baselines}
The first simple baseline algorithm (denoted as \texttt{baseline1}) is to treat the current number of purchases as the future number of  purchases, and hence it guarantees less than 100\% relative error, given that the number is increasing and always positive.

Another baseline algorithm (denoted as \texttt{baseline2}) is to assume a linear relationship between the current number of purchases and the future number of purchases. Suppose we know the number of purchases $N_{t_1}$ at time $t_1$, and aim to predict the number of purchases $N_{t_2}$ at time $t_2$, where $t_1 < t_2$. Then we assume that
\begin{equation}
N_{t_2} = \alpha N_{t_1} +\beta
\end{equation} 
where $\alpha$ and $\beta$ is model parameters that can be learned from training data.

\subsubsection{Social Propagation Model}
As seen in Figure~\ref{fig:growth-prediction}, the growth in sales after tipping in Groupon is described well by a multiplicative process. What follows from the model is that to 
obtain the popularity for the next time step we multiply the current popularity by a small, random amount. More specifically, let $t_1$ and $t_2$ denote two different time steps and $t_1 < t_2$. Following \cite{SzaboH10CACM}, we have
\begin{equation}
\log N_{t_2} \approx \log(N_{t_1}) + \sum_{t=t_1}^{t_2}r(t)X_t
\end{equation}
according to Equation~(\ref{eqn:approximate})

This process, called ``growth with random multiplicative noise", describes the dynamics of users' attention to web contents~\cite{FangB07PNAS}. While the increments at each time step are random, their expected value over many time steps adds up ultimately to $\sum_{t=t_1}r(t)X_t$ in the log-linear model, where $\sum_{t=t_1}r(t)X_t$ accounts for the linear relationship between the log-transformed popularities at different times $t_1$ and $t_2$.

\begin{figure}[htb]
  \centering
  \subfigure[Groupon] {\includegraphics[width=1.5in]{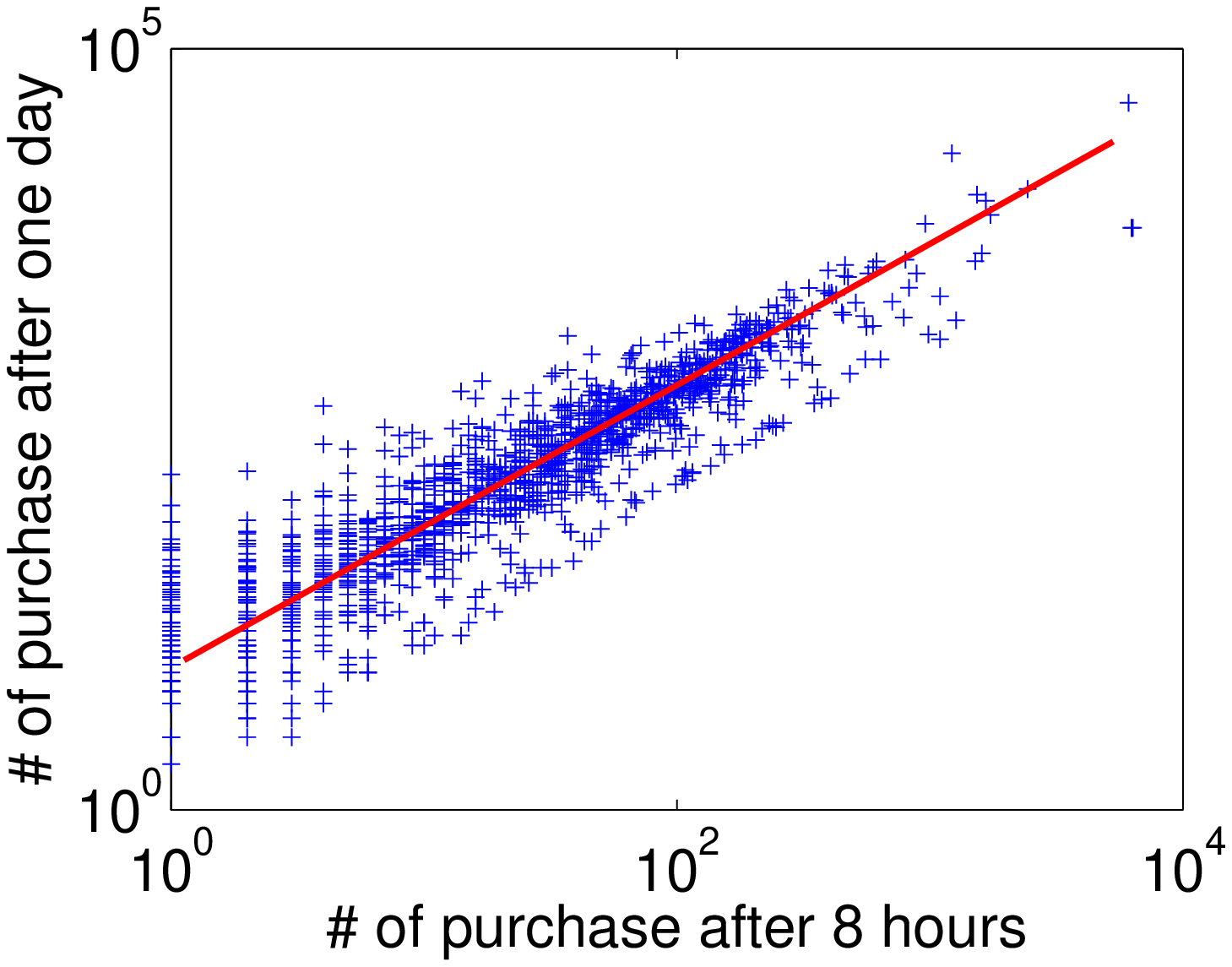}}  
\subfigure[LivingSocial]{
    \includegraphics[width=1.5in]{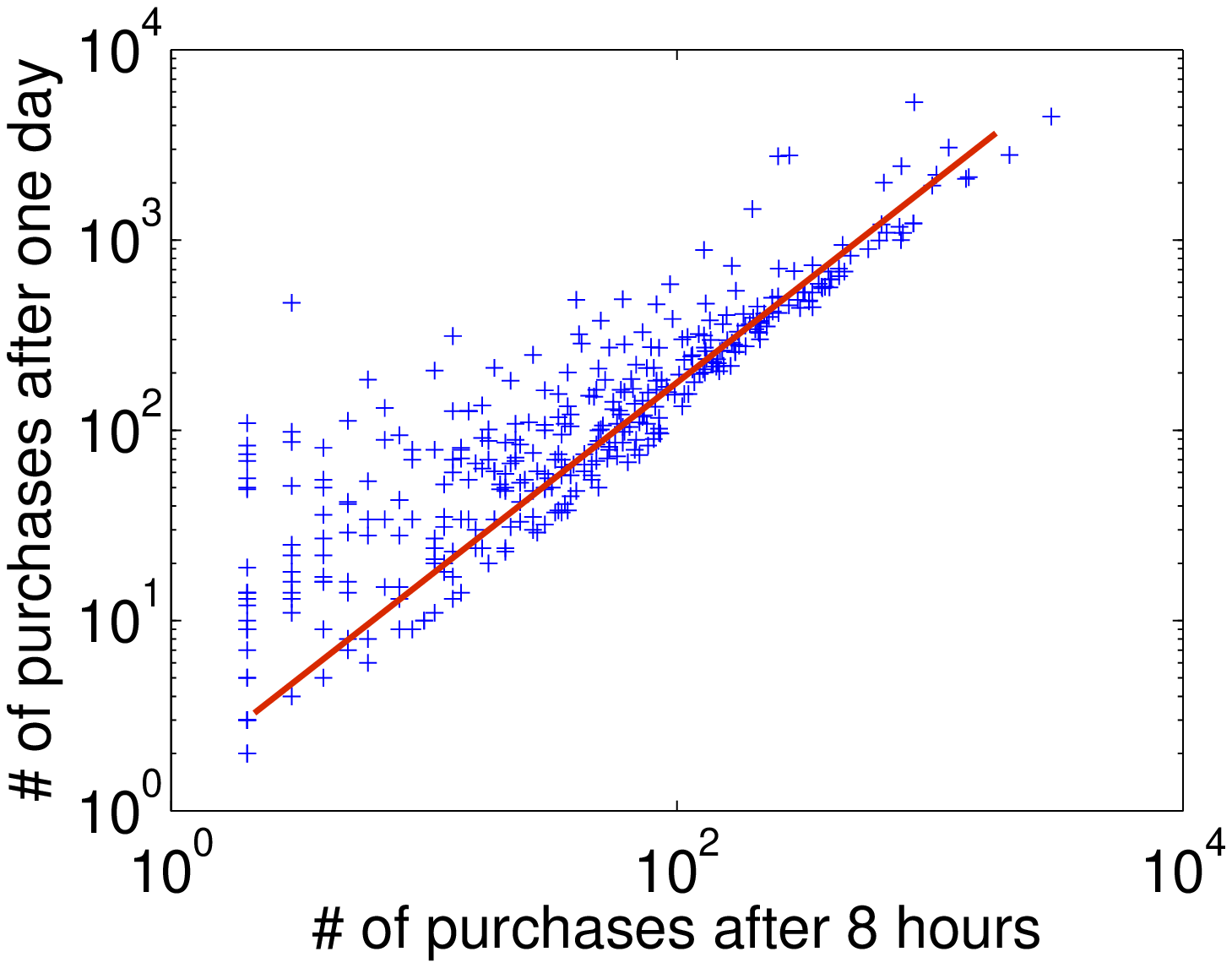}}
 \caption{Number of purchases after 8 hours vs. number of purchases after one day (log-scale). The bold line is the linear fit to the data}\label{fig:prediction_loglogtest}
\end{figure}

Here, we introduce the process used to model and predict the future number of purchases of a deal. We first perform a logarithmic transformation on the number of purchases, similar to \cite{SzaboH10CACM,Byers11CoRR}. To help determine whether the number of purchases early on is a predictor of later number of purchases, see Figure~\ref{fig:prediction_loglogtest}, which shows the number of purchases at the reference time $t_1  = 8$ hours vs. the number of purchases at the end of a day (i.e., $t_2 = 24$ hours) in both Groupon and LivingSocial. We logarithmically rescaled the horizontal and vertical axes in the figure to show
the number of purchases for different deals, which span four orders of magnitude.

Figure~(\ref{fig:prediction_loglogtest}) shows that there is a strong correlation between the earlier observations of the number of purchases of a deal and the later observations. So we can determine the linear regression coefficients between $t_1$ and $t_2$ on a given training dataset, and then use the estimated coefficients to extrapolate on the test dataset.

\begin{figure*}[htb]
  \centering
 \subfigure[Baseline-1] {\includegraphics[width=1.9in]{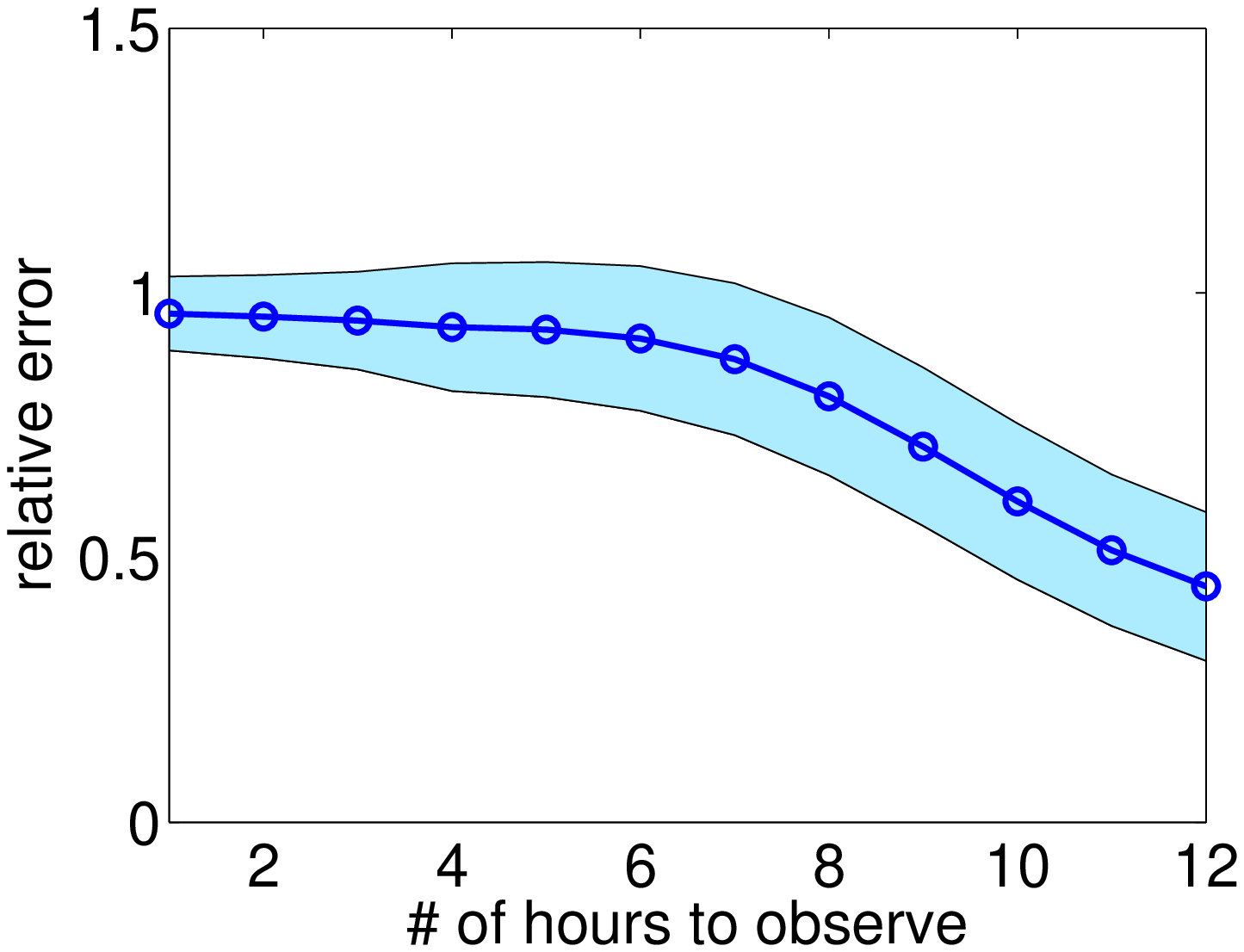}}   
\subfigure[Baseline-2] {\includegraphics[width=1.9in]{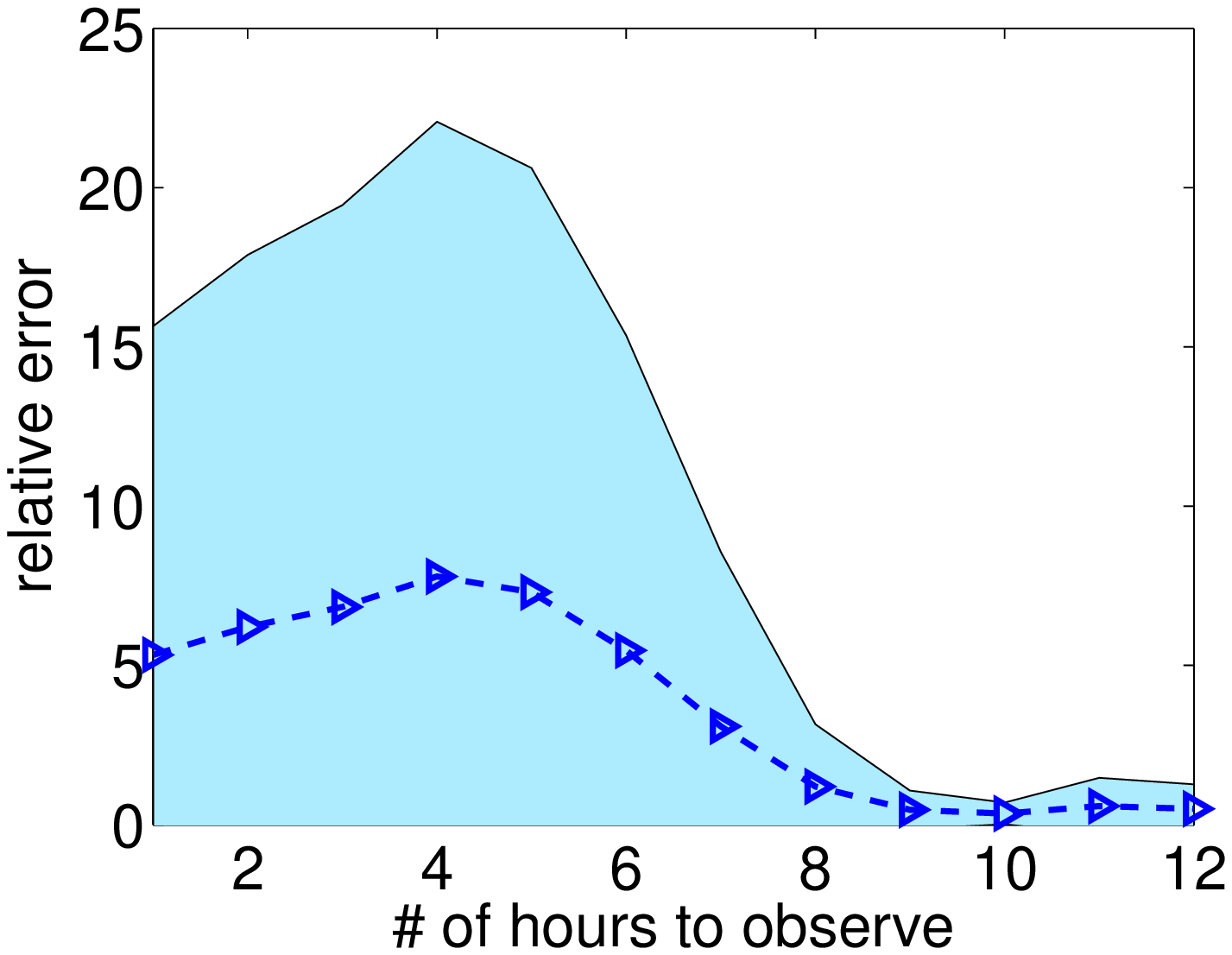}}  
  \subfigure[Multi-Linear Regression (MLR) Model]{
    \includegraphics[width=1.9in]{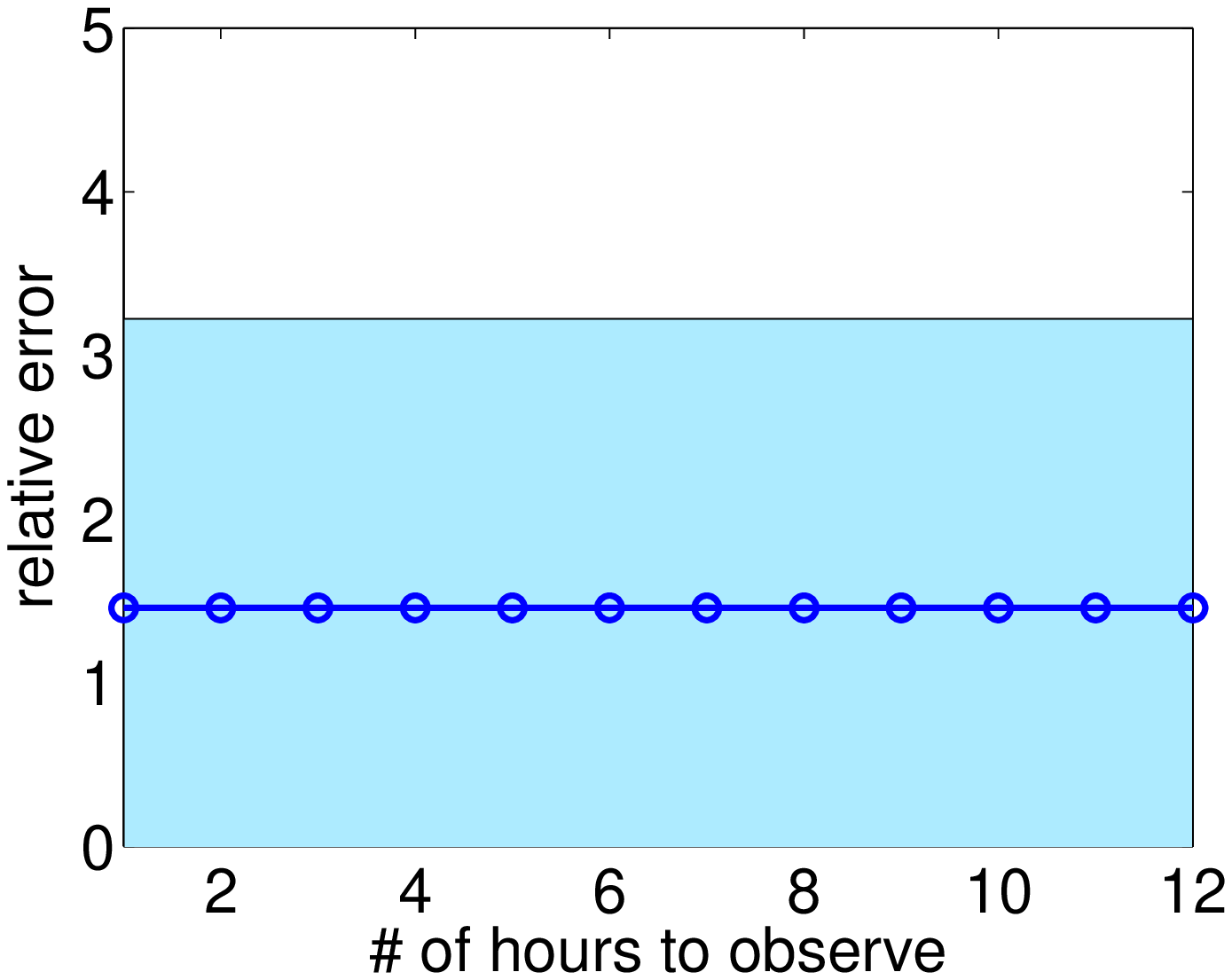}} \\
  \subfigure[Social Propagation (SP) Model]{
    \includegraphics[width=1.9in]{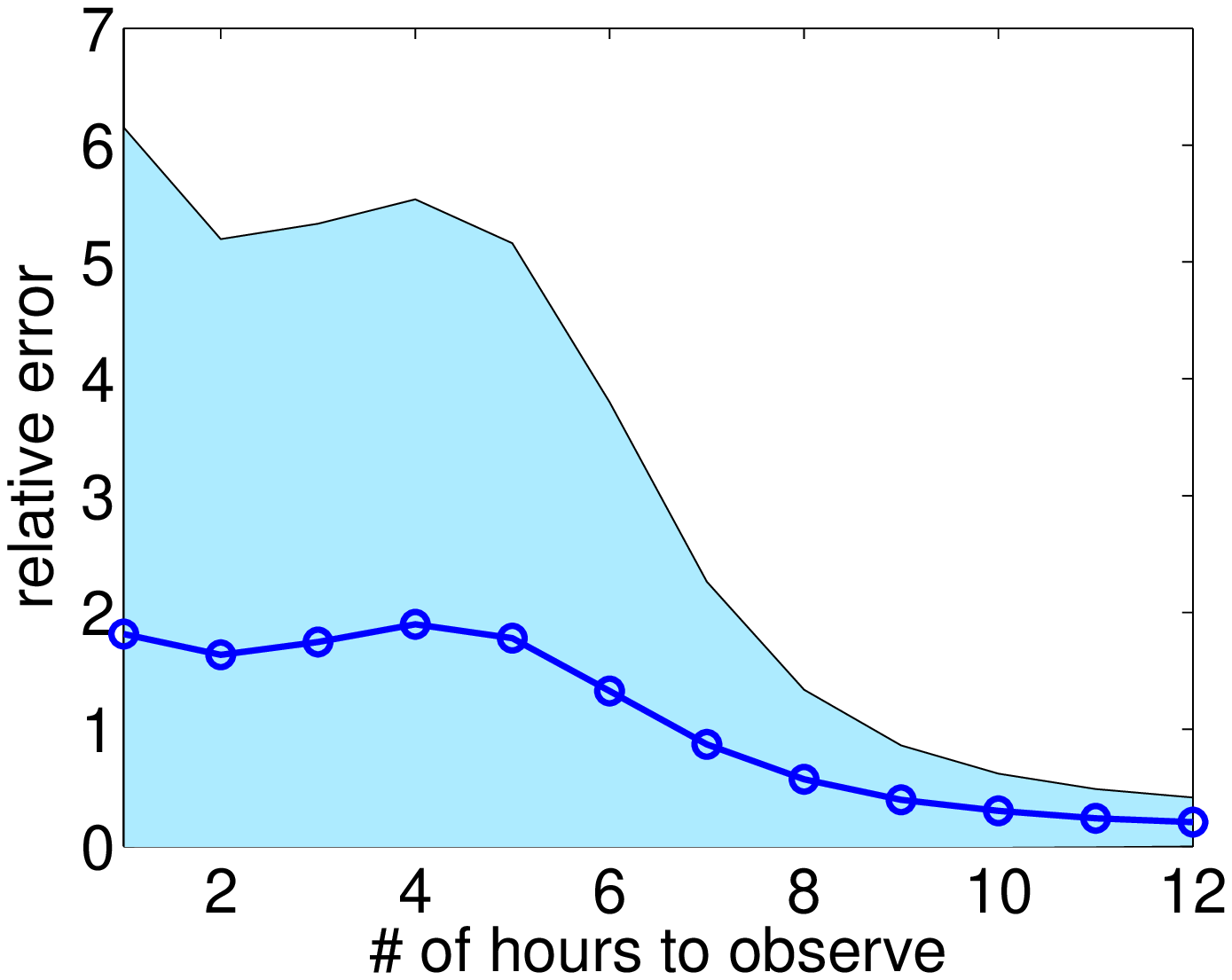}} 
\subfigure[Comparison]{\includegraphics[width=1.9in]{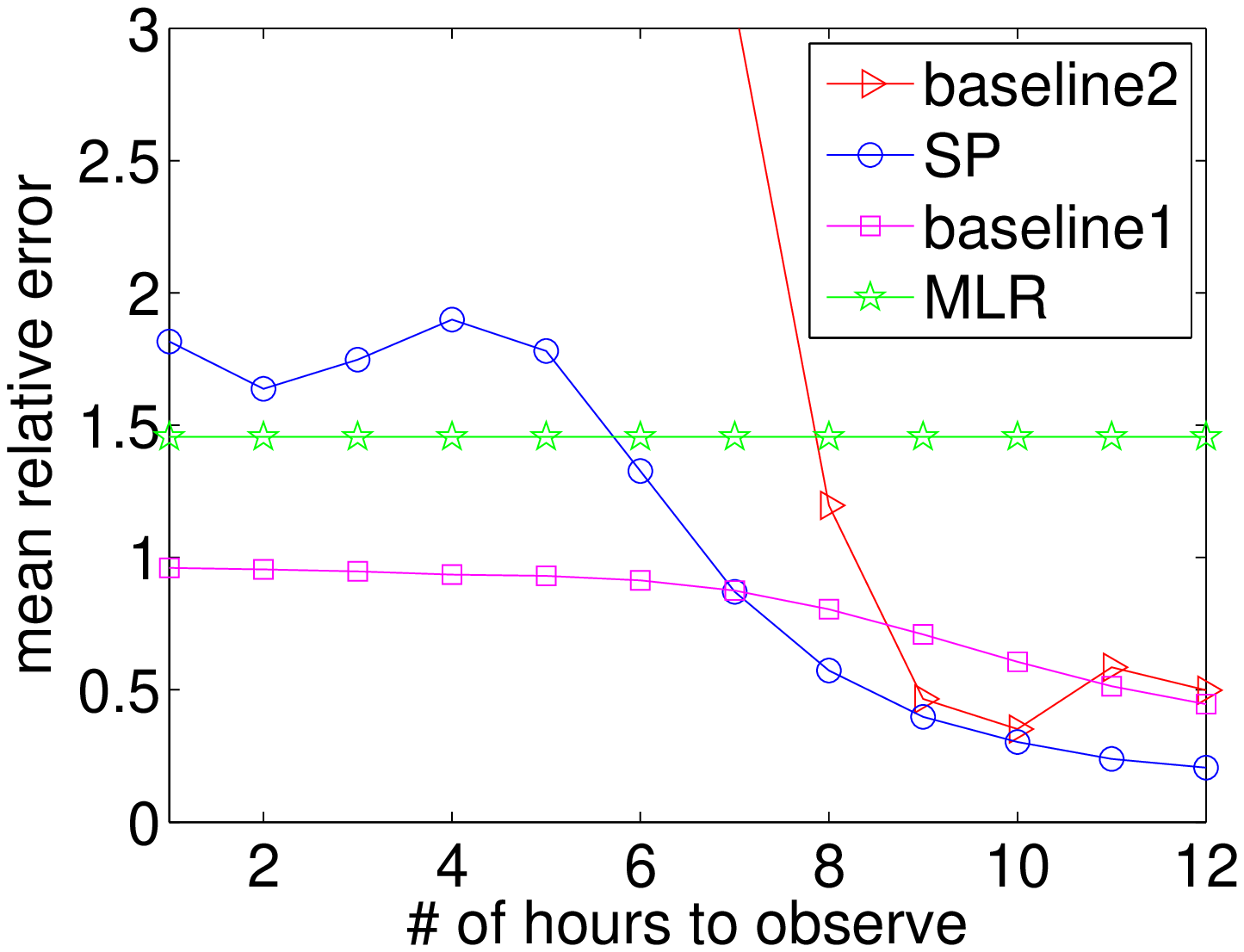}} 
\subfigure[Relative error distribution]{\includegraphics[width=1.9in]{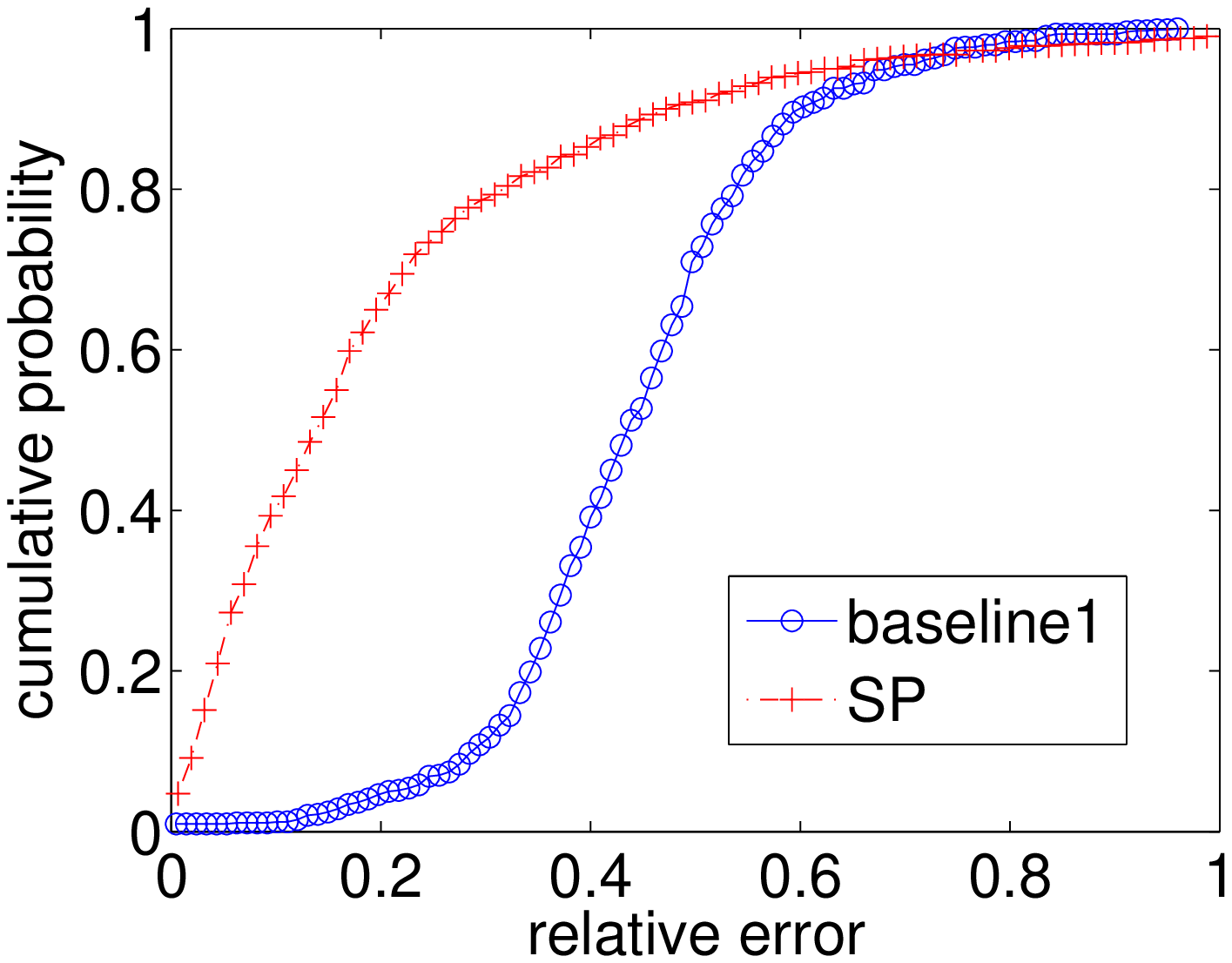}} 
 \caption{Performance comparison of prediction of the number of purchases after one day in Groupon. In (a)-(e), lines denote the average relative error, and shaded regions cover the areas of one-standard error.}\label{fig:prediction_evaluation_G}
\end{figure*}

Note that there is a limitation to this approach. As we discussed before, in Groupon a renewal process, rather than a multiplicative one, governs the dynamics before tipping. So this approach may not perform well for the very early observations. Nevertheless, it is applicable to both Groupon and LivingSocial since the multiplicative process is the main process during the life cycle of a deal for both services.

%\subsection{Without previous observations}
%Since it is always desirable to make accurate predictions as early as possible we now discuss a method that does not require previous observations. 
%
%Even without any previous observations of how many people have bought the deal in very early hours, we know some basic information about a deal, such as whether it is in featured position, retail price, discount etc.. Furthermore, we know the tipping point if it is a Groupon deal. As mentioned before, those attributes of deals can help estimate the number of purchases, so we now utilize a multi-linear regression model to help the prediction task.
%
%Let $y$ denote the number of purchases of a deal after one day. We then have
%\begin{equation}
%y = \beta_0 + \beta_1 a_1+ \beta_2 a_2 + \cdots 
%\end{equation}  
%where $a_i (i = 1, 2, \cdots)$ denotes the attributes of a deal such as tipping point, discount and so on, and $\beta_j (j = 0, 1, 2, \cdots)$ are the coefficients of this model. 
%
%With a given training dataset, we can estimate optimal settings for those coefficients, i.e., $\beta_j^*$ , and then for a given test deal with attributes $a_1$, $a_2$, $\cdots$ ,we predict that  the number of purchase of this deal after one day $\hat{y}$ is
%\begin{equation}
%\hat{y} = \beta_0^* + \beta_1^* a_1 + \beta_2^* a_2 + \cdots
%\end{equation}  
%
%
\subsection{Evaluation}
In this subsection, we conduct an  experimental study to evaluate the proposed prediction algorithms. As  discussed before, the important task is to be able to predict how successful a deal will be.  Since there are many deals with a lifetime of one day we evaluate the performance of different algorithms by how accurately they can predict the number of purchases of a deal after one day. Here, we use relative error, i.e., $\frac{|\mbox{real purchases - predicted purchases}|}{\mbox{real purchases}}$,  as the performance metric to measure accuracy.   

\begin{table*}%[h]
\centering
\begin{tabular}{|l|c|c|c|}
\hline
\multicolumn{4}{|p{17cm}|}{[Deal Title: The Magnetic Field - Asheville] \$12 for Two Tickets to a Theater Performance (Up to \$28 Value)} \\\hline
Algorithms & Real purchases & Predicted purchases &  Relative error\\
\hline
\texttt{baseline-1} 12-hour observation& 251 & 93  & 0.63\\ \hline
\texttt{baseline-2} 12-hour observation& 251 & 482 & 0.92\\ \hline
\texttt{MLR} & 251 & 51  & 0.80\\ \hline
\texttt{SP} with 12-hour observation & 251 & 355 & 0.42\\ \hline
\hline
\hline
\multicolumn{4}{|p{17cm}|}{[Deal Title: Lime Leaf Thai Cuisine - Hendersonville] \$10 for \$20 Worth of Thai Fusion Cuisine} \\\hline
Algorithms & Real purchases & Predicted purchases & Relative error \\
\hline
\texttt{baseline-1} 12-hour observation& 384 & 169  & 0.56\\ \hline
\texttt{baseline-2} 12-hour observation& 384 & 714 & 0.86\\ \hline
\texttt{MLR} & 384 & 1,452  & 2.783\\ \hline
\texttt{SP} with 12-hour observation & 384 & 463 & 0.21\\ \hline
\hline
\end{tabular}
\caption{Example prediction results for Groupon deals. }\label{tbl:evaluation-G}
\end{table*}

\begin{figure*}[htb]
  \centering
\subfigure[Baseline-1] {\includegraphics[width=1.9in]{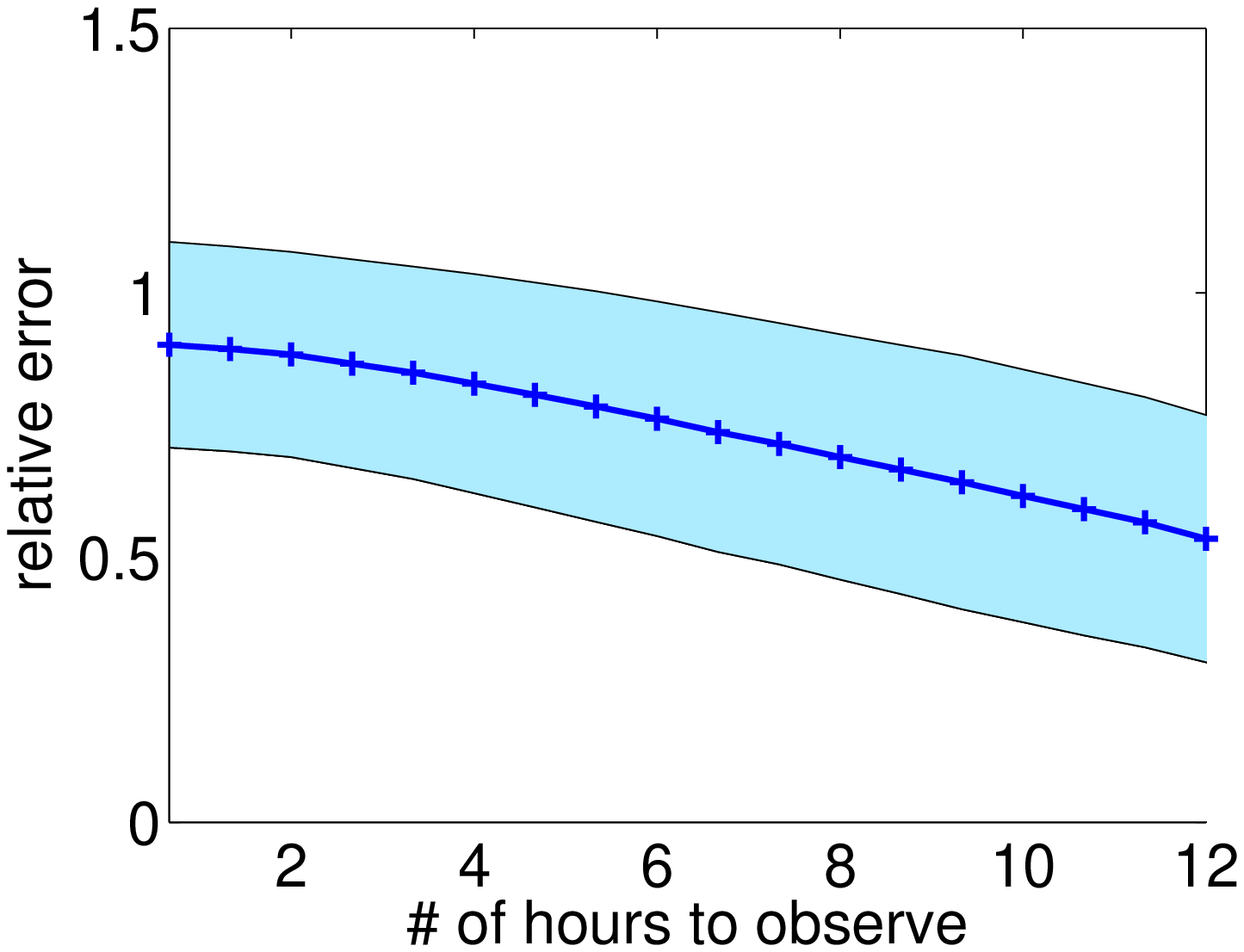}}
  \subfigure[Baseline-2] {\includegraphics[width=1.9in]{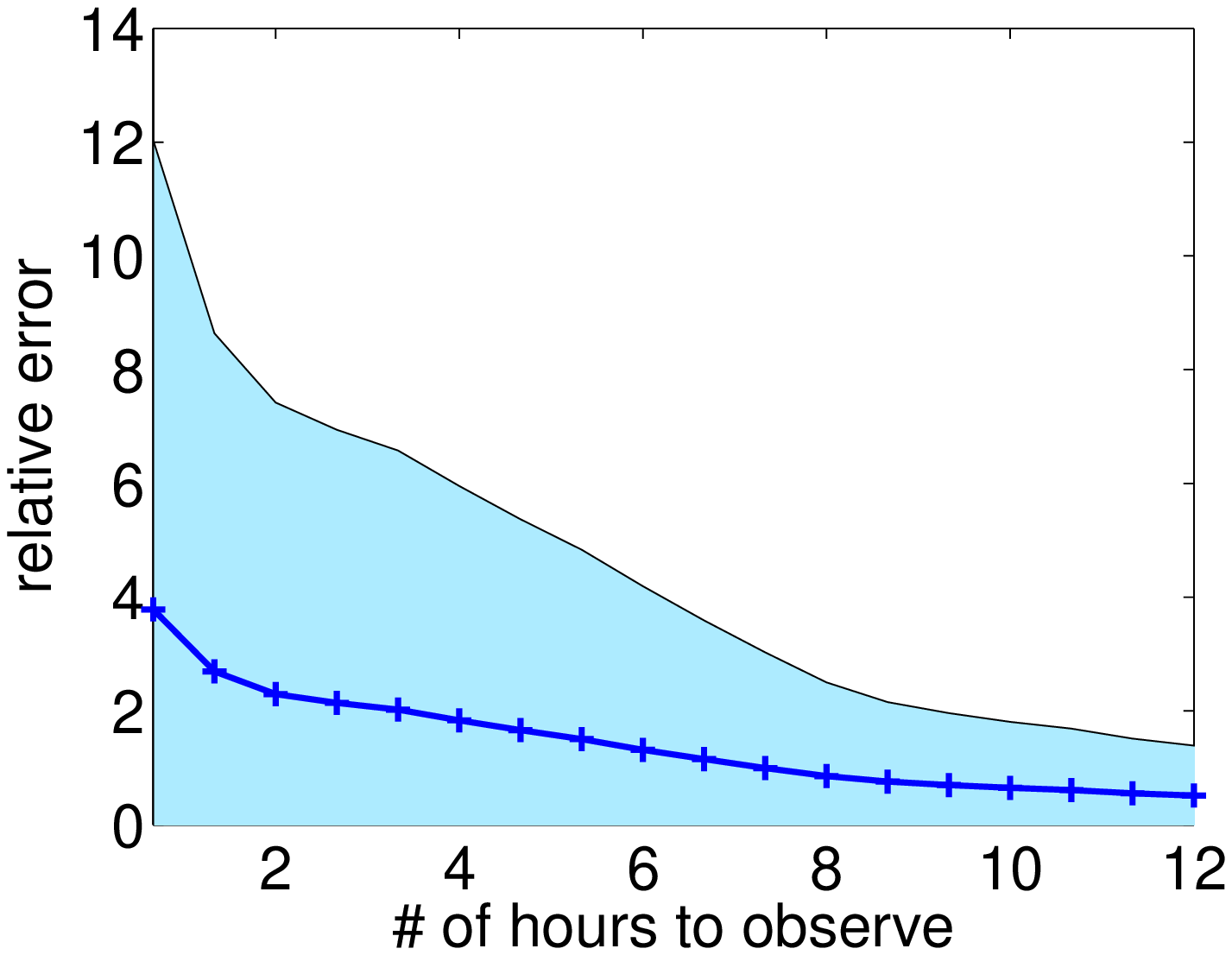}} 
  \subfigure[Social Propagation Model]{
    \includegraphics[width=1.9in]{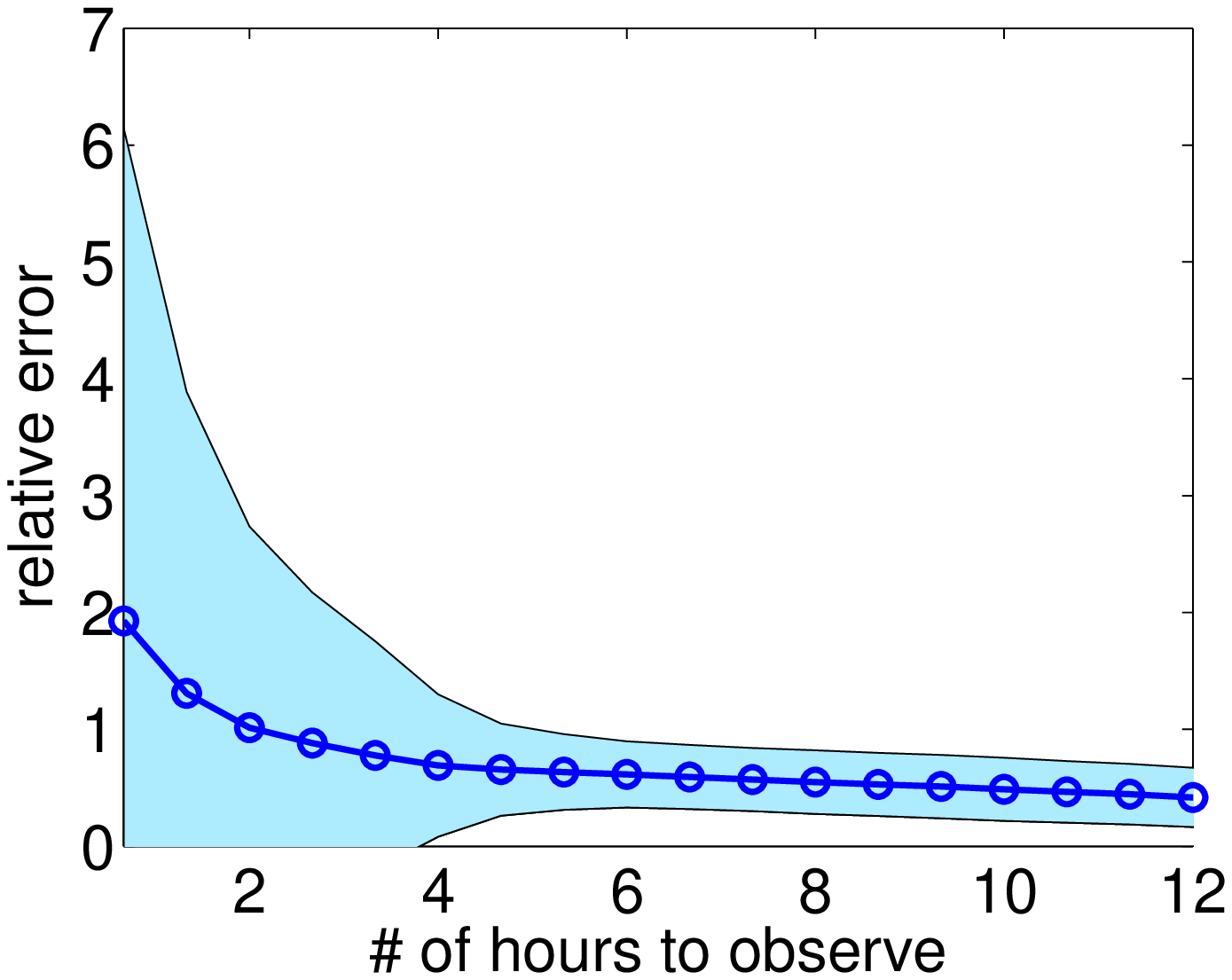}}\\ 
\subfigure[Comparison]{\includegraphics[width=1.9in]{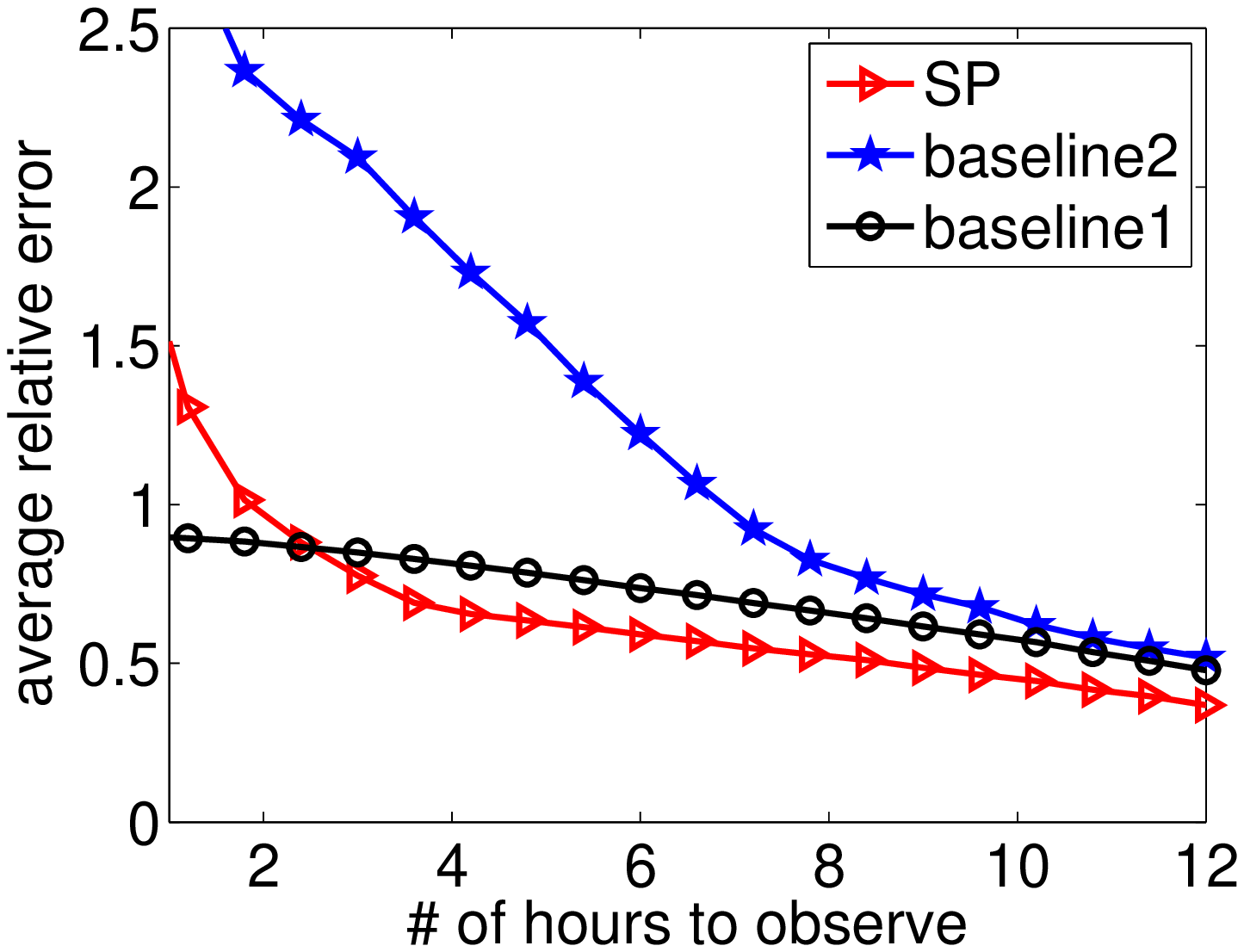}} 
\subfigure[Relative error distribution]{\includegraphics[width=1.9in]{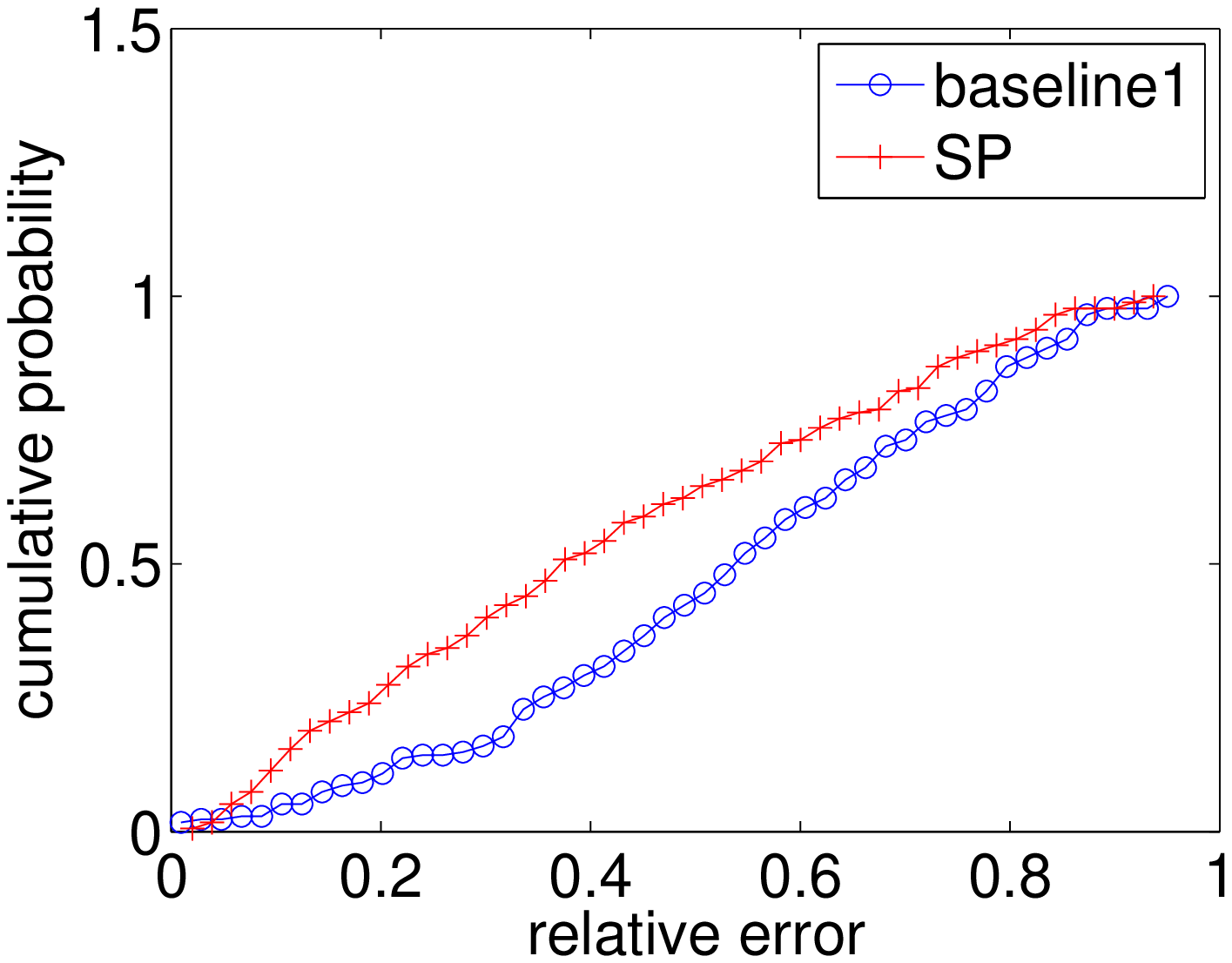}} 
 \caption{Performance comparison of prediction of the number of purchases after one day in LivingSocial.In (a)-(d), lines denote the average relative error, and shaded regions cover the areas of one-standard error.}\label{fig:prediction_evaluation_L}
\end{figure*}
%\begin{figure}[htl]
%\centering
%
%\caption{MRSE and MRE distributions for the case of prediction based on social propagation with 12-hour observation in Groupon}\label{fig:groupon-error-distribution}
%\end{figure}
\subsubsection{Experiments with Groupon Deals}
First, we conduct experiments on the Groupon dataset by randomly splitting it into halves, where one half is used for training and another half is for testing.

In Figure~\ref{fig:prediction_evaluation_G}, we find \texttt{baseline1} shows the best performance 
among all the testing algorithms with less than 7-hours of observations. After 7-hour observation, our proposed social propagation model (denoted as \texttt{SP}) shows the best performance. Note that a deal which attracts more than hundred purchases within the first hour after launching (6 deals in total in the experiment) is treated differently by applying \texttt{baseline1}, as these deals are extremely popular and don't follow the general multiplicative process. The justification for applying \texttt{baseline1} is that, these deals are so appealing that local merchants usually place quantity limits. 

As we observed before, deals in Groupon are usually tipped after  
about 7 hours. Before tipping, the purchase dynamics is governed by random discovery instead of the multiplicative process, thus the social propagation model fails to achieve good performance. However, we find that there is an inflection point which occurs at about 7 hours. After 7 hours of observations, 
the social propagation model exhibits relatively good performance, and it performs much better with more hours of observation.  In Figure~\ref{fig:prediction_evaluation_G} (f), relative error distributions of \texttt{baseline1} and \texttt{SP} with 12-hour observation are examined. We find that the relative error is less than 50\% for over 90\% of deals when using \texttt{SP}, and there are about 70\% of deals achieving less than 20\% relative error when applying \texttt{SP}. %relative error in about 0.5 probability, and more importantly, deals with accuracy of $0.1\sim0.4$ with \texttt{SP} are in much larger amount than the ones with \texttt{baseline1}. Note that in Figure~\ref{fig:purchase_growth}(a), it takes time (around 7  hours) for groupon deals to be tipped and enter into the multiplicative process phase. Therefore,  it only experiences about 5-hour multiplicative process for a deal, if we predict the one day purchases with previous 12-hour observation for the deal. To allow three more hours observation, we find that there are over 60\% of deals achieving less than 0.5 relative error by using \texttt{SP} as shown in Figure~\ref{fig:prediction_evaluation_G}(g).

In the experiment, we incorporated all the attributes of the deals into the multi-linear regression (denoted as \texttt{MLR}) model,  including the tipping point. Tipping points can be considered as the observation of the number of purchases at around 6-8 hours. Therefore, as shown in Figure~\ref{fig:prediction_evaluation_G}(f), the multi-linear regression model achieves a comparable  performance with our model within an observation period of 6 hours.  
To exemplify the prediction accuracy, we show the results from a few Groupon deals in Table~\ref{tbl:evaluation-G}. 

%both  have much smaller squared relative error than bass diffusion model does. Nevertheless, we need to point out that the prediction results from social propagation model and multi-linear regression model  

As a refinement for Groupon deals, we perform \texttt{baseline1} if the deal has not tipped; otherwise, we apply the social propagation (\texttt{SP}) model.

\begin{table*}%[h]
\centering
\begin{tabular}{|l|c|c|c|}
\hline
\multicolumn{4}{|p{17cm}|}{[Deal Title: Coastal Contacts] \$60 to Spend on Prescription Eyeglasses (Now \$19)} \\\hline
Model & Real purchases & Predicted purchases & Relative error\\
\hline
\texttt{baseline1} with 12-hour observation & 129 & 32 & 0.75\\ \hline
\texttt{baseline2} with 12-hour observation& 129 & 245 & 0.90 \\ \hline
\texttt{SP} with 12-hour observation & 129 & 110  & 0.14\\ \hline
\hline
\hline
\multicolumn{4}{|p{15cm}|}{[Deal Title: Dawgs!] \$10 (Pay \$5) or \$20 (Pay \$10) to Spend on Food and Drink} \\\hline
Model & Real purchases & Predicted purchases & Relative error\\
\hline
\texttt{baseline1} with 12-hour observation & 75 & 28 & 0.63 \\ \hline
\texttt{baseline2} with 12-hour observation& 75 & 147 & 0.96 \\ \hline
\texttt{SP} with 12-hour observation & 75 & 110  & 0.47\\ \hline
\hline
\end{tabular} 
\caption{Example prediction results for LivingSocial deals.}\label{tbl:evaluation-L}
\end{table*}
\subsubsection{Experiments with LivingSocial Deals}
We conducted similar experiments on the LivingSocial dataset. As shown in Figure~\ref{fig:prediction_evaluation_L}, our social propagation model (\texttt{SP}) always outperforms \texttt{baseline2} and beats \texttt{baseline1} with more than 2-hours of observations. 
Because of the limitations of the crawling technique, we do not have  information about which deal is the featured one in a given city; and there is no tipping point in LivingSocial, which prevents the multi-linear regression model from generating good predictions. However, the social propagation model shows very good performance in LivingSocial. In particular, we examine the distribution of relative errors for predictions based on \texttt{SP} and \texttt{baseline1} with 12-hours of observations in LivingSocial. As shown in Figure~\ref{fig:prediction_evaluation_L}(e), we find that there are about 65\% of deals with less than 50\% relative error; and \texttt{SP} always outperforms \texttt{baseline1}.

Similarly, we show prediction results from some LivingSocial deals in Table~\ref{tbl:evaluation-L}. As shown in Table~\ref{tbl:evaluation-L}, the social propagation model exhibits better prediction performance than both baselines, in terms of relative error. 

Finally, our design for purchase prediction of Groupon deals is that we perform \texttt{baseline1} if with less than 3-hour observation; otherwise, we apply the social propagation (\texttt{SP}) model.
Note that due to different mechanisms in Groupon and LivingSoical, inflection points are placed at very different times (i.e., 6-8 hours in Groupon, and 2-4 hours in LivingSocial). Therefore, \texttt{SP} can be applied earlier in LivingSocial than in Groupon. However, as shown in Figure~\ref{fig:prediction_evaluation_G}(e) and  Figure~\ref{fig:prediction_evaluation_L}(e), the relative error measured on the test set decreases rapidly for Groupon, while for LivingSocial the prediction converges more slowly to the actual value. After 17 hours, the expected relative error obtained when estimating one-day purchases of a deal by using \texttt{SP} is about 20\%, while the same relative error is attained 13 hours after a Groupon deal is launched.   This is due to the fact that novelty decay is faster in Groupon than in LivingSocial, i.e. it takes 7 hours in Groupon to reach the saturating point; while it takes about 14 hours in LivingSocial to reach the saturating point in Figure~\ref{fig:growth-prediction}. So it is easier to predict the one-day purchases of Groupon deals with fewer hours of observations (after tipping). One possible explanation of this is that the tipping point incentive mechanism for propagating deals in Groupon disappears after the tipping point has been
reached. In LivingSocial, on the other hand, the incentive to propagate a deal is always present for at least some
users and furthermore the individual gain of propagating is greater.

\section{Conclusions}\label{sec:conclusion}
%In this paper, we studied the collective attention and purchase dynamics of group deals.
%We used Groupon as a case study to empirically verify our models.
%While we concentrated on data from Groupon, we stress that our model is general enough so as to capture the
%purchase behaviors in group deals with both random discovery and social propagation behaviors.
%In future work we intend to extend our model to capture purchase cancelation behavior,
%multi-day dynamics, as well as decay factor dependency on tipping point. Furthermore,
%we plan to conduct baseline, and state-of-the-art benchmark prediction tests
%of a predictor of final purchase volume derived from our model.
%We are also interested in studying additional group deal sites to see how social
%propagation mixes with random discovery in settings where there may not be any
%clear inflection point such as the tipping point in Groupon.
In this paper, we presented a study of the group purchasing behavior
of daily deals in Groupon and LivingSocial and introduced a predictive dynamic model of collective
attention for group buying behavior. Using large data sets from both Groupon  and LivingSocial we showed how the model 
was able to predict the popularity of group deals as a function of time.
Our main finding is that the different incentive mechanisms in Groupon and LivingSocial lead to different
propagation behavior, which in turn leads to differences in predictability. 
However, the basic stochastic processes as well as the distributional parameters
of growth and decay are strikingly similar. Given that Groupon no longer provides detailed statistics
of purchases over time, the models presented here can not easily be applied by any observer. However,
both deal site owners and merchants should be able to benefit from analyzing the early stream of purchases
using the models presented here. Our work also gives some insights into how different incentive mechanisms
can affect the longevity of propagation momentum. These insights could be exploited in local marketing campaigns
where viral and social dissemination of offers is desirable. 
%\begin{acknowledgments}
%
%\end{acknowledgments}

\bibliographystyle{abbrv}
%\bibliography{reference}

\end{document}